\documentclass[aps,pre,twocolumn,reprint,showpacs,superscriptaddress,preprintnumbers,amsmath,amssymb,floatfix,raggedbottom]{revtex4-1}
\usepackage{graphicx}  
\graphicspath{{figures/}}
\usepackage{natbib}
\usepackage[breaklinks=true,colorlinks=true,linkcolor=blue,citecolor=blue,urlcolor=blue,hypertexnames=false,bookmarks=false]{hyperref} 
\usepackage[all]{hypcap} 
\usepackage{dcolumn}   
\usepackage{bm}        
\usepackage{amsmath}   
\usepackage{amssymb}   
\usepackage{amsfonts}
\usepackage{float}
\usepackage{color}
\usepackage{bm}
\usepackage{latexsym}
\usepackage{epsfig}
\usepackage{listings}
\usepackage{comment}
\usepackage{wasysym}
\usepackage{mathtools}
\usepackage[utf8]{inputenc}
\usepackage[T1]{fontenc}
\usepackage{silence}
\begin{document}


\title[]{Vortex ring induced large bubble entrainment during drop impact}
\author{Marie-Jean \surname{Thoraval}}
\affiliation{
Division of Physical Sciences and Engineering $\&$ Clean Combustion Research Center,
King Abdullah University of Science and Technology (KAUST),
Thuwal, 23955-6900, Saudi Arabia
}
\affiliation{
Physics of Fluids Group, Faculty of Science and Technology, Mesa+ Institute,
University of Twente, 7500 AE Enschede, The Netherlands
}
\affiliation{
International Center for Applied Mechanics, State Key Laboratory for Strength and Vibration of Mechanical Structures,
Xi'an Jiaotong University, Xi'an, 710049, People's Republic of China
}
\author{Yangfan \surname{Li}}
\affiliation{
Mechanical Engineering, National University of Singapore,
9 Engineering Drive 1, Singapore 117576
}
\author{Sigurdur T. \surname{Thoroddsen}}
\affiliation{
Division of Physical Sciences and Engineering $\&$ Clean Combustion Research Center,
King Abdullah University of Science and Technology (KAUST),
Thuwal, 23955-6900, Saudi Arabia
}
\date{\today}

\begin{abstract}

For a limited set of impact conditions, a drop impacting onto a pool can entrap an air bubble as large as its own size.
The subsequent rise and rupture of this large bubble plays an important role in aerosol formation
and gas transport at the air-sea interface.
The large bubble is formed when the impact crater closes up near the pool surface and is known to occur
only for drops which are prolate at impact.
Herein we use experiments and numerical simulations to show that a concentrated vortex ring,
produced in the neck between the drop and pool, controls the crater deformations and pinch-off.
However, it is not the strongest vortex rings which are responsible for the large bubbles,
as they interact too strongly with the pool surface and self-destruct.
Rather, it is somewhat weaker vortices which can deform the deeper craters, which manage to pinch off the large bubbles.
These observations also explain why the strongest and most penetrating vortex rings emerging from drop impacts, are not produced by oblate drops
but by more prolate drop shapes, as had been observed in previous experiments.
\end{abstract}

\maketitle

\section{Introduction}

When a liquid drop impacts onto a liquid pool, it can entrap air bubbles of different sizes
\cite{Esmailizadeh1986, Prosperetti1989, Oguz1989, Oguz1990, Prosperetti1993, Rein1993, Thoroddsen2003, Smith2003, Korobkin2008, Lee2011, Saylor2012, Thoroddsen2012, Ray2012, Tran2013, Semenov2013, Ray2015, Semenov2015, Beilharz2015, Murphy2015}.
While the smallest bubbles can dissolve, contributing to the gas transfer into the pool \cite{Esmailizadeh1986},
the largest bubbles rise to the surface where they burst.
The latter mechanism can produce many smaller bubbles \cite{Bird2010} and aerosols in the atmosphere \cite{Lee2011, Lhuissier2012, Wilson2015}.
Different mechanisms have been identified, responsible for air entrapment \cite{Liow2007}.
The recent progress in high-speed video imaging has allowed researchers to explain some aspects of bubble entrapment \cite{Thoroddsen2012},
and even discover new mechanisms \cite{Thoraval2012, Castrejon-Pita2012, Thoraval2013}.

Here we focus on the so-called \textit{large bubble entrainment} identified by Pumphrey and Elmore \cite{Pumphrey1990}.
They observed  that for a small range of water drop diameters and impact velocities,
the cavity produced by the drop expands radially below the pool surface and closes at the top to entrap a large bubble.
They already suggested that the drop shape oscillations play an important role in this entrapment.
Zou \textit{et al.} \cite{Zou2012} have recently reported a similar large bubble entrapment for horizontally restricted pools.
They determined that such bubbles were observed only for prolate drops, i.e. those with a larger vertical extent.
Finally, Wang, Kuan and Tsai \cite{Wang2013} have shown that the large bubble entrapment on unrestricted pools also occurs only for prolate drops,
and over a wider range of conditions than originally observed \cite{Pumphrey1990}.

However, the underlying mechanism for this large bubble entrapment has remained a mystery until now.
We combine herein high-speed imaging and numerical simulations to show
the crucial role played by vorticity in deforming the interface and entrapping the large bubble.
The role of vorticity in this process has been ignored in the above studies.
It is indeed only recently that the effects of vorticity on drop-impact splashing have been demonstrated
\cite{Thoraval2012, Castrejon-Pita2012, Thoraval2013, Agbaglah2015, Lee2015}.

\section{Experimental setup}

\begin{figure}
 \begin{center}
  \includegraphics[width=\linewidth]{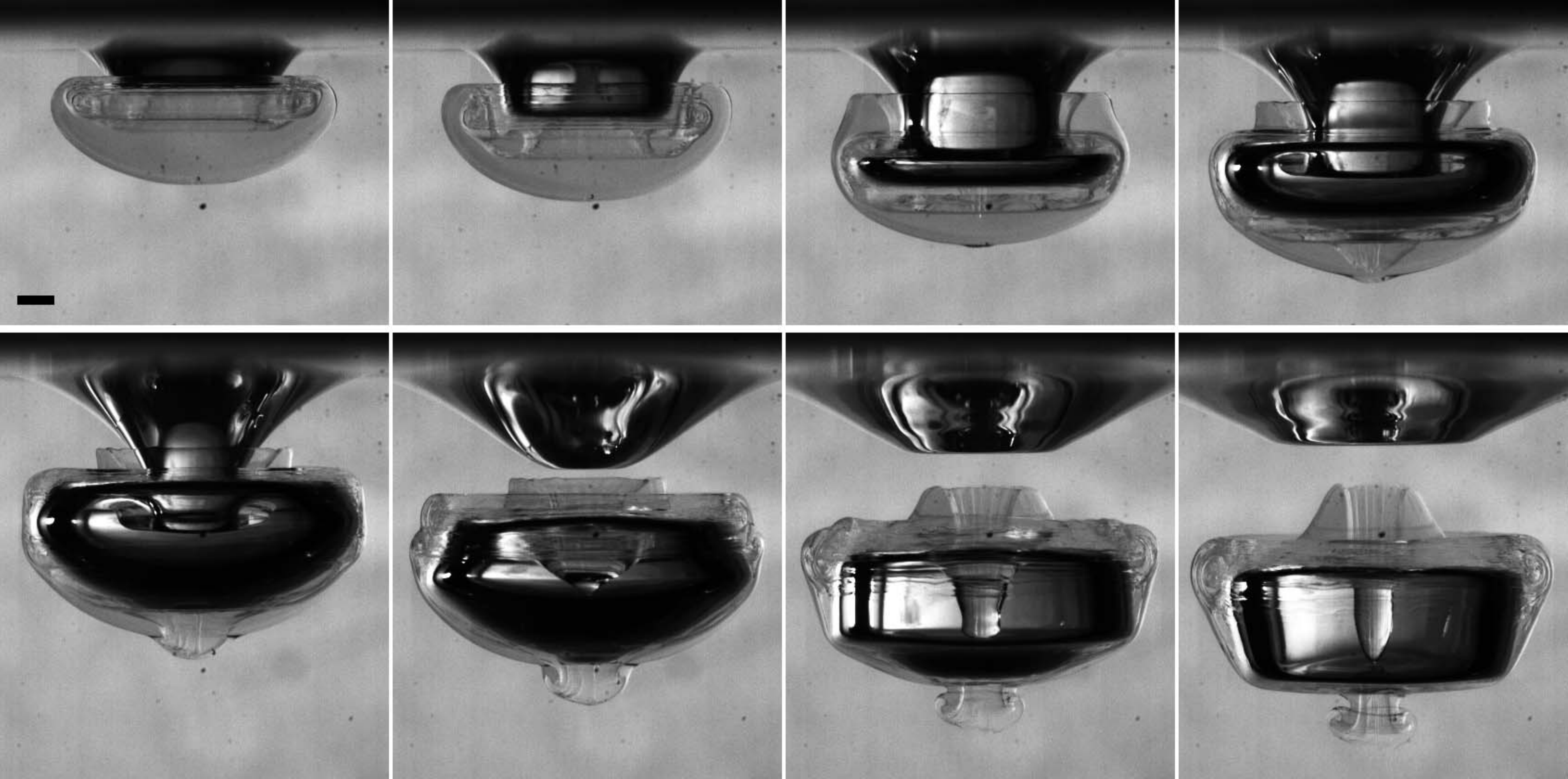}
  \caption{Side view imaging of a prolate drop impacting into a pool ($D =$~5.0~mm, $V =$~0.9~m/s),
	observed below the pool surface	at the following times after the first contact: 7.7, 8.7, 11, 13, 16, 20, 24, 27~ms.
	A slightly different liquid is used in the drop (10.5\%~$MgSO_{4}$ solution with viscosity 1.96~cP and density 1.105~g/cm$^3$)
	and the pool (distilled water) to visualize their interface (hemispherical and curled lined in the first frame).
	The dark region is the air cavity created by the impacting drop.
	The scale bar is 1~mm long.}
  \label{fig:LBexperiments}
 \end{center}
\end{figure}

We use drops pinched off from a nozzle and impacting on a pool surface.
We adjust the impact velocity by varying the impact height $H$.
By using slightly different liquids in the drop and the pool, the difference in refractive index
reveals the deformation of the interface between the two liquids,
while keeping the main features of the flow unchanged (see Section~\ref{sec:Other}).
Fig.~\ref{fig:LBexperiments} shows such a case. The first image shows the presence of a strong vortex ring,
made visible by the roll-up of the liquid-liquid interface.
The vorticity is produced by the flow around the curved free surface during the early dynamics \cite[p. 366]{Batchelor1967},
and has been the subject of many studies \cite{Thomson1885, Chapman1967, Rodriguez1988, Rein1993, Peck1994, Cresswell1995, Dooley1997, Peck1998b, Josserand2003, Saylor2004, Watanabe2008, Li2008, Santini2013, Moore2014},
as will be discussed in Section~\ref{sec:discussion}.
In the following images, the air-liquid interface is pulled around and towards the core of the vortex,
first radially outwards in the third frame, then upwards in the fourth frame.
This creates a tongue of liquid above the cavity, pulled towards the center by surface tension,
which finally closes to entrap the large bubble.
The pulling on the interface can be explained by the lower pressure at the core of the vortex and the velocity field around it \cite{Yu1990, Jha2015}.

\section{Numerical simulations}

We perform numerical simulations to understand the effect of the different parameters on the vorticity dynamics.
We use the \textit{Gerris} code \cite{Popinet2003, Popinet2009},
validated in previous studies on drop impact \cite{Thoraval2012, Thoraval2013}.
It solves the Navier-Stokes equations with a \textit{Volume-of-Fluid} method for the two-phase flow and a dynamic adaptive grid refinement.

We simulate the axisymmetric impact of a water drop onto a water pool initially at rest, including gravity.
A passive tracer is initially placed into the drop (colored in red) to identify the origin of the water,
in a similar way as in the experiment in Fig.~\ref{fig:LBexperiments}.
The main tracer used to follow the air-water interface is not affected by this additional tracer.

The use of numerical simulations allows a precise control of the drop shape, independently of the impact velocity.
We compare here the impact of a spherical drop to ellipsoids of revolution.
The equivalent drop diameter is defined as $D = \left(D_h^2 D_v\right)^{1/3}$,
with horizontal diameter $D_h$ ($a = D_h/D$) and vertical diameter $D_v$ ($b = D_v/D$).
The vertical diameter is obtained by volume conservation: $b=  1/a^2$.
The simulation starts before the drop touches the pool,
with a uniform vertical drop impact velocity $V$.

In the axisymmetric domain of size 1, the drop has a radius of 0.13 and the pool depth is 2.5$D$.
The mesh is refined near the air/water interface, the passive tracer interface
and in the regions of high vorticity, at a maximum refinement level of 11,
corresponding to the smallest cells of size $2^{-11}$ ($ D / \delta x = 532$).
Some numerical details are given in Thoraval~\textit{et al.} \cite{Thoraval2012}.
Only at the upper boundary of the parameter regime do we see some effect of increasing resolution (see Supplemental Material \cite{SM}).
This would benefit from even higher level of mesh refinement but is prohibitive due to the long duration of the entrapment process,
especially compared to the first ejecta studied in \cite{Thoraval2012}, were such super-resolution simulations are feasible.

All * variables are non-dimensionalized with $D$ and $V$, including $t^{*} = t / \tau$, with $\tau = D/V$.

\section{Results}

\begin{figure*}
 \begin{center}
  \includegraphics[width=5.25in]{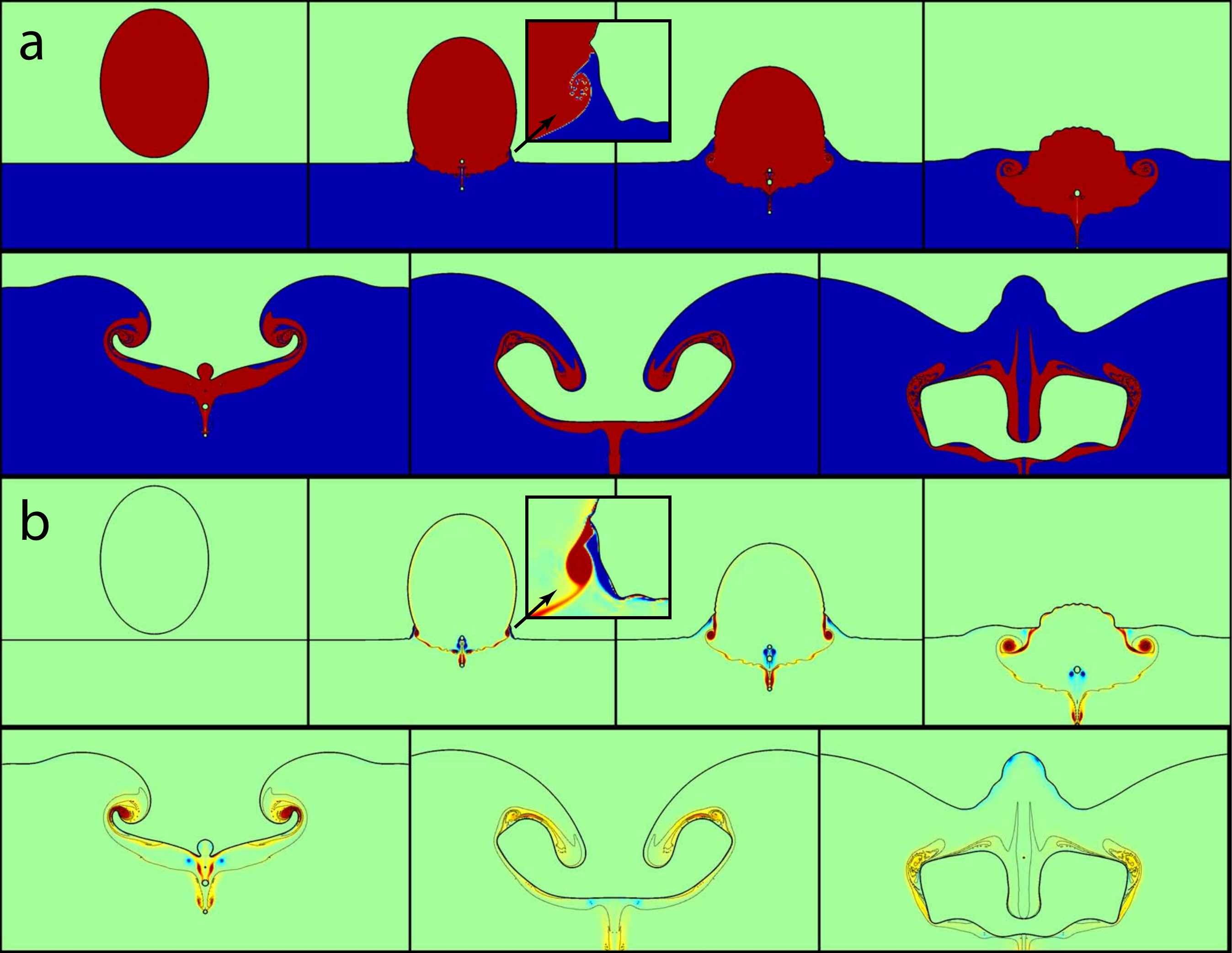}
  \caption{Typical time sequence leading to large bubble entrapment,
	showing the following times: $t^{*} = $~-0.05, 0.18, 0.41, 0.87, 1.80, 2.99, 4.80.
	$D =$~5~mm, $V =$~1.0~m/s, $a =$~0.9 ($Re=$~5000, $We=$~69).
	(a) The water originally in the drop (resp. pool) is colored in red (blue), while the air is in green.
	The liquids in the drop and the pool are identical, without any material interface between them.
	(b) The corresponding vorticity field.}
  \label{fig:largeBubble}
 \end{center}
\end{figure*}

Fig.~\ref{fig:largeBubble} shows a typical time evolution leading to the large bubble entrapment,
for a prolate drop under conditions similar to Fig.~\ref{fig:LBexperiments}.
Despite the use of different liquids in the experiments, and the uncertainty on the drop shape,
the simulation captures very well the dynamics,
including the rectangular shape of the bubble after pinch-off, and the vertical jet in the center.
We see clearly the early shedding of vorticity on the side of the drop (inset of Fig.~\ref{fig:largeBubble}b),
hidden in experiments because of the curved interface above the pool surface, but visible by x-rays \cite{Agbaglah2015}.
The side vortex ring then rolls-up with the trailing vortex sheet into a larger vortex ring,
that pulls radially on the interface below the surface of the pool (see Supplemental Videos \cite{SM}).

\subsection{Role of gravity}

\begin{figure}[h!]
 \begin{center}
  \includegraphics[width=\linewidth]{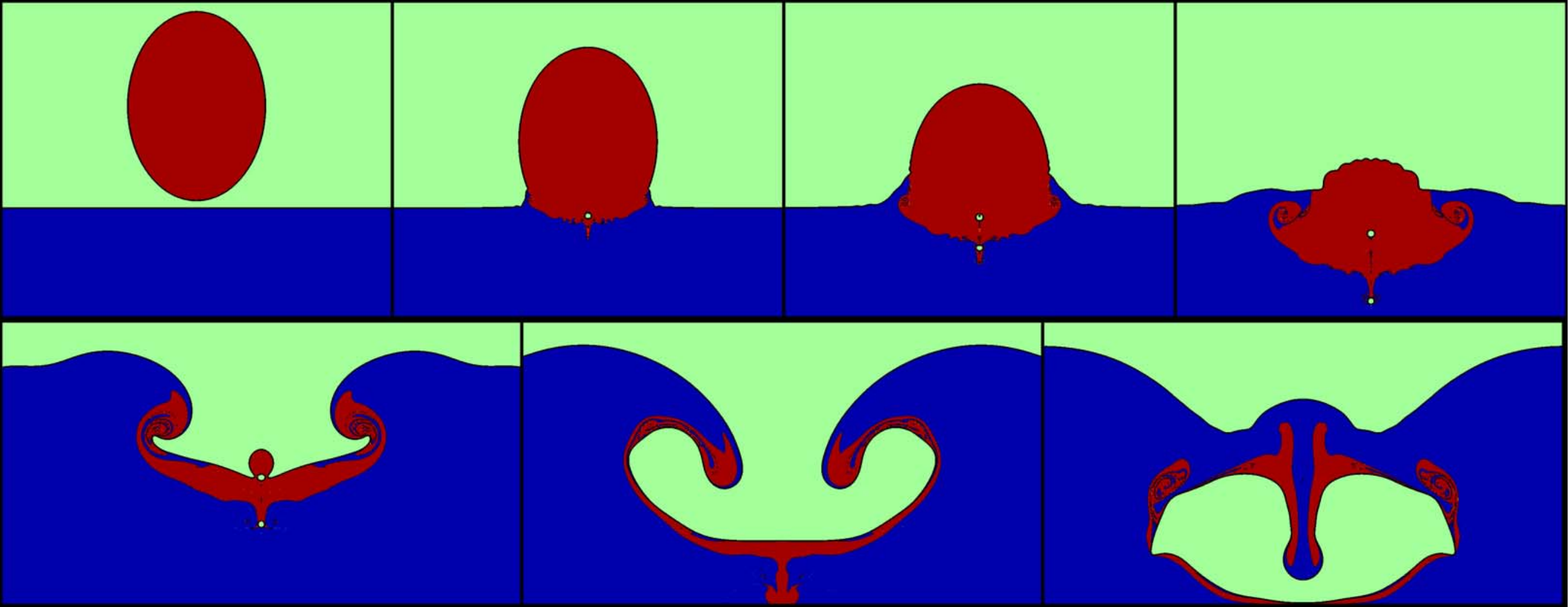}
  \caption{Influence of gravity on the large-bubble entrapment.
  This sequence is for the same conditions as in Fig.~\ref{fig:largeBubble}, which includes gravity,
	while in this figure gravity has been excluded.
	$t^{*} = $~-0.05, 0.18, 0.41, 0.87, 1.80, 2.99, 4.80.
	$D =$~5~mm, $V =$~1.0~m/s, $a =$~0.9.}
  \label{fig:largeBubble-nog}
 \end{center}
\end{figure}

What is the role of gravity in this entrapment process?
This can be directly addressed in the numerical simulations, where we can simply turn gravity off.
The typical Froude number $Fr = V/\sqrt{gD}$ is rather low ($Fr = $~4.5 in Fig.~\ref{fig:largeBubble}).
Gravity can therefore be neglected during the initial impact of the drop ($t^{*} \leq 1$),
but may affect the dynamics at later times. This is indeed what is observed.
Figure~\ref{fig:largeBubble-nog} shows that the large bubble entrapment still occurs, even if gravity is removed.
The sequences show no significant differences up to the time of entrainment of the bubble.
Only in the last image, at time $t^{*} = $~4.8, does gravity influence the bottom shape of the bubble.
Gravity is therefore not responsible for the large bubble entrapment.

\begin{figure}[h!]
 \begin{center}
	\includegraphics[width=\linewidth]{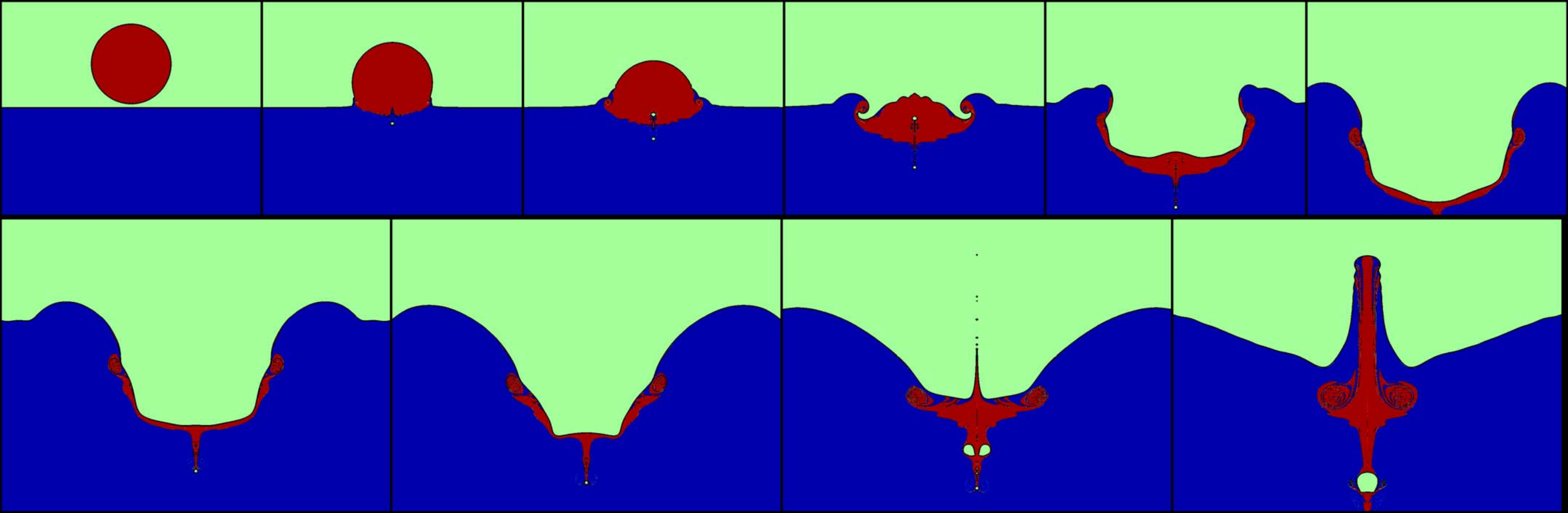}\vspace{1.5mm}\\
  \includegraphics[width=\linewidth]{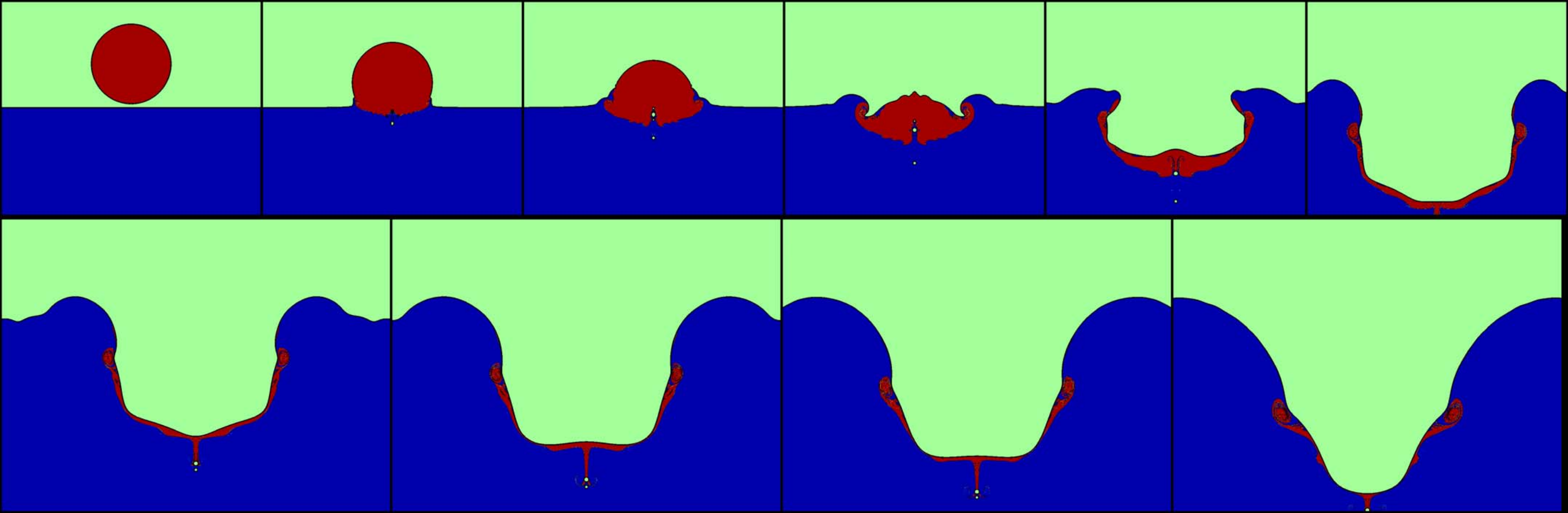}
  \caption{Role of gravity on regular bubble entrapment, at the bottom of the collapsing crater.
  Same impact conditions as in Fig.~\ref{fig:largeBubble}, but now for a spherical drop, $a =$~1.
	With gravity (top) or without (bottom).
	$t^{*} = $~-0.05, 0.18, 0.41, 0.87, 1.80, 2.99, 3.64, 4.80, 5.72, 7.64.
	$D =$~5~mm, $V =$~1.0~m/s.}
  \label{fig:regularEntrainment}
 \end{center}
\end{figure}

In contrast, gravity plays a key role in the regular bubble entrainment,
which is produced at the bottom of the collapsing impact crater \cite{Prosperetti1989, Oguz1990, Chen2014}
and is greatly affected by the hydrostatic pressure.
This was also verified with our simulations as shown in Fig.~\ref{fig:regularEntrainment}.
Regular entrainment, by pinch-off of the bottom of the crater, is no longer observed if gravity is removed.

This difference is consistent with the different time scales of these entrainment mechanisms.
The large bubble entrapment occurs at $t^{*} \sim 3$ (see Figs.~\ref{fig:largeBubble}, \ref{fig:entrapmentTZ}a),
while the regular entrapment is observed at longer times
($t^{*} \simeq 5.6$ in our simulations, or 9.4 in \cite[their Fig.~4]{Pumphrey1990}).

\subsection{Parameter space}

\begin{figure}
 \begin{center}
  \includegraphics[width=0.8\linewidth]{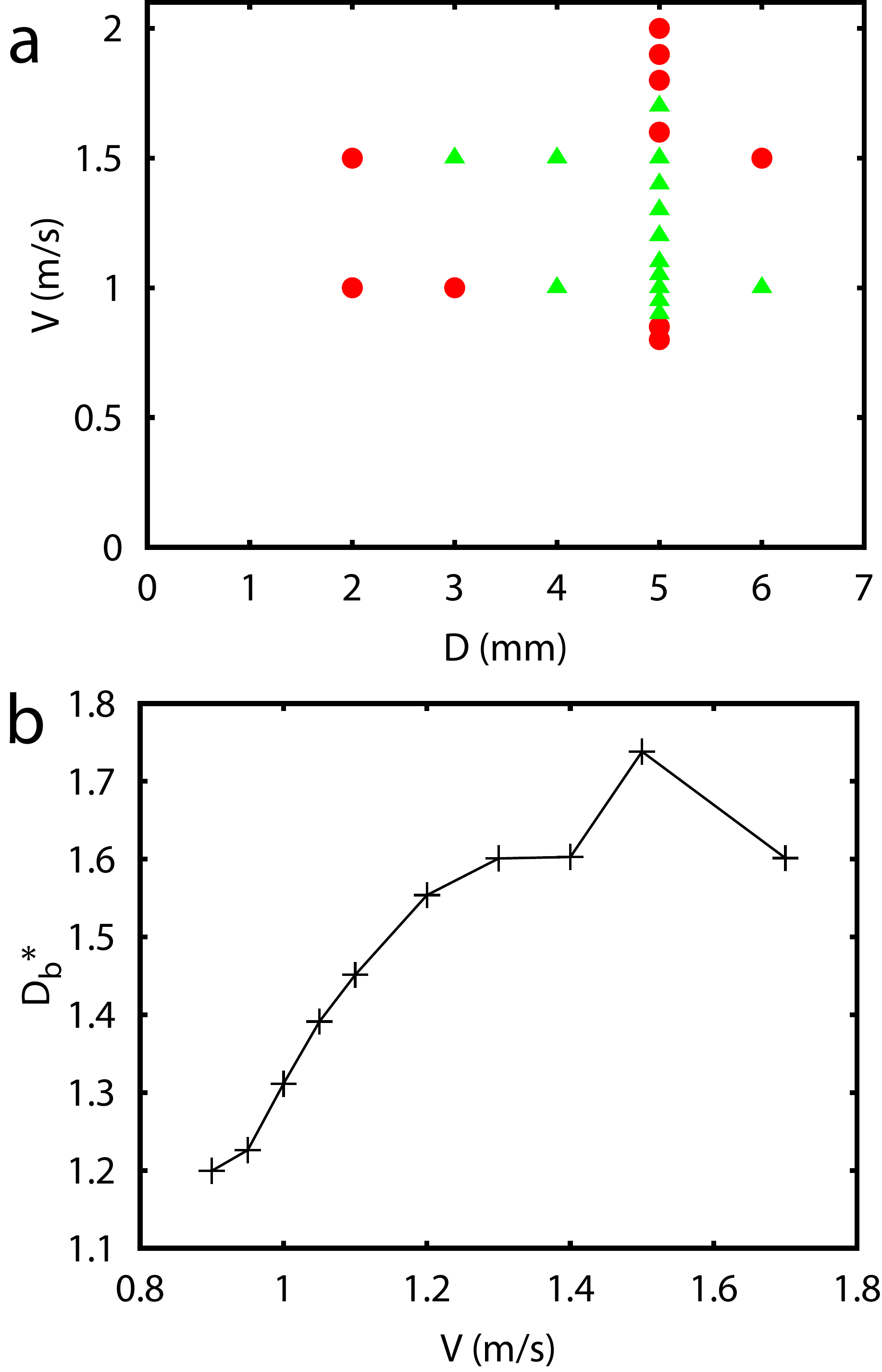}
  \caption{(a) Parameter space where a large bubble is entrapped (\textcolor{green}{$\blacktriangle$}) or not (\textcolor{red}{$\CIRCLE$}).
	All simulations are for a prolate water drop with horizontal axis $a= D_h/D = 0.9$.
	No large bubble entrapment was observed for other drop aspect ratios when $a$ in 0.95, 1, 1.05 and 1.1.
	(b) Equivalent spherical diameter of the large air bubble $D_b$,
	based on its total volume and normalized by the drop diameter ($D_b^* = D_b/D$), for $D =$~5~mm and $a =$~0.9.}
  \label{fig:entrapment}
 \end{center}
\end{figure}

We explore the effect of the impact velocity $V$, drop diameter $D$ and drop shape on the large bubble entrapment.
Fig.~\ref{fig:entrapment}(a) shows the $D$-$V$ parameter space where large bubbles are entrapped (\textcolor{green}{$\blacktriangle$}).
For all cases in the figure, we have simulated 5 different drop shapes
with horizontal diameter $a = D_h/D$ values of 0.9, 0.95, 1, 1.05 and 1.1.
Large bubble entrapment occurs only for the largest vertical elongation of the drop (the most prolate), i.e. $a = 0.9$,
in the parameter range reported in Fig.~\ref{fig:entrapment}(a).
The goal of this study is to understand the underlying mechanism of the large bubble entrapment,
rather than mapping an extensive parameter space.
By comparing here drops with the same volume and impact velocity,
we are able to confirm the crucial effect of the drop shape, as previously suggested.
The parameter space is consistent with the experiments of Wang \textit{et al.} \cite{Wang2013},
showing similar lower and upper limits on both axis.
The less clear upper limit in impact velocity ($V \simeq 1.6$~m/s) is probably due to insufficient numerical refinement
for such high Reynolds number ($Re = \rho D V / \mu = 8000$),
as was demonstrated by the very high refinement needed in the study of Thoraval \textit{et al.} \cite{Thoraval2012},
when capturing the details of the earliest contact.
Keep in mind that our regime of (\textcolor{green}{$\blacktriangle$}) is continuous,
whereas in experiments, the prolate phase of the pinched-off oscillating drop
is restricted to certain discrete release heights and thereby impact velocities \cite{Wang2013, Thoraval2013}.

The size of the large bubble increases with the impact velocity (Fig.~\ref{fig:entrapment}b),
except for the last point, which lies at the limit of the entrapment region.
The bubble diameter is 20 to 75\% larger than the drop.
For the highest impact velocities these simulation numbers are still sensitive to refinement level.
This trend is consistent with the experiments of Zou~\textit{et al.} \cite{Zou2012}, although their study was on a horizontally confined pool.

\begin{figure}
 \begin{center}
	\includegraphics[width=0.8\linewidth]{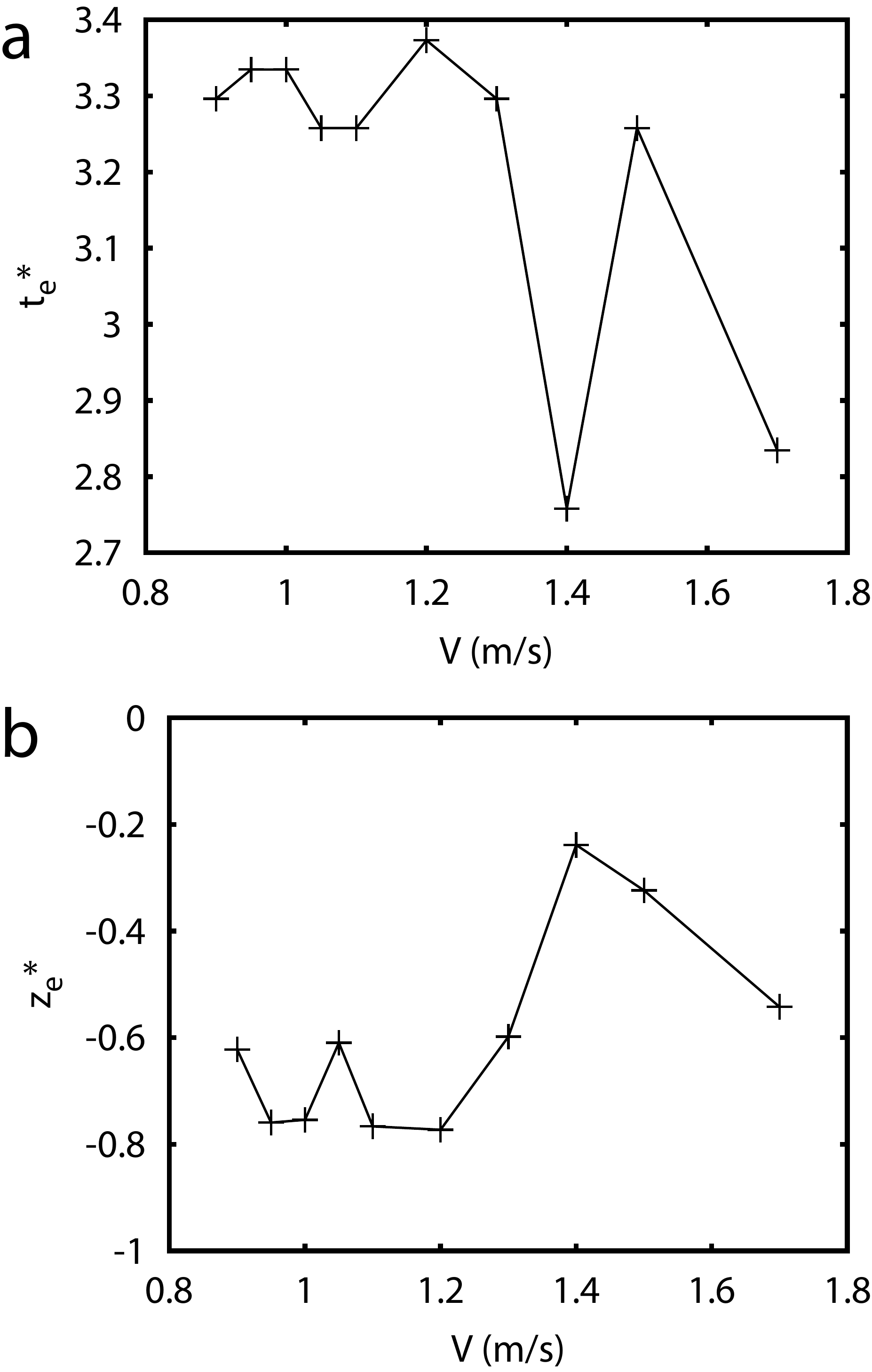}
  \caption{Time (a) and depth (b) of the large bubble entrapment.
	It is defined as the point where the liquid tongue first touches the opposite symmetrical liquid tongue
	on the axis of symmetry above the air cavity.
	The zero in depth (b) is set at the initial pool surface.}
  \label{fig:entrapmentTZ}
 \end{center}
\end{figure}

The top enclosure and entrapment occurs in most cases at $t^{*} \simeq 3.3$
with a couple of the higher-velocity impacts closing in shorter time $\simeq 2.8$ (Fig.~\ref{fig:entrapmentTZ}a).
This is consistent with Fig.~\ref{fig:LBexperiments} ($t^{*} \simeq 3.2$),
and previous observations ($t^{*} \simeq 3.5$ in \cite{Pumphrey1990} and 2.7 in \cite{Zou2012}).
The bubble closes below the original pool surface, at about one drop radius in depth (Fig.~\ref{fig:entrapmentTZ}b).

\subsection{Vortex pulling}

\begin{figure*}
 \begin{center}
  \includegraphics[width=0.8\linewidth]{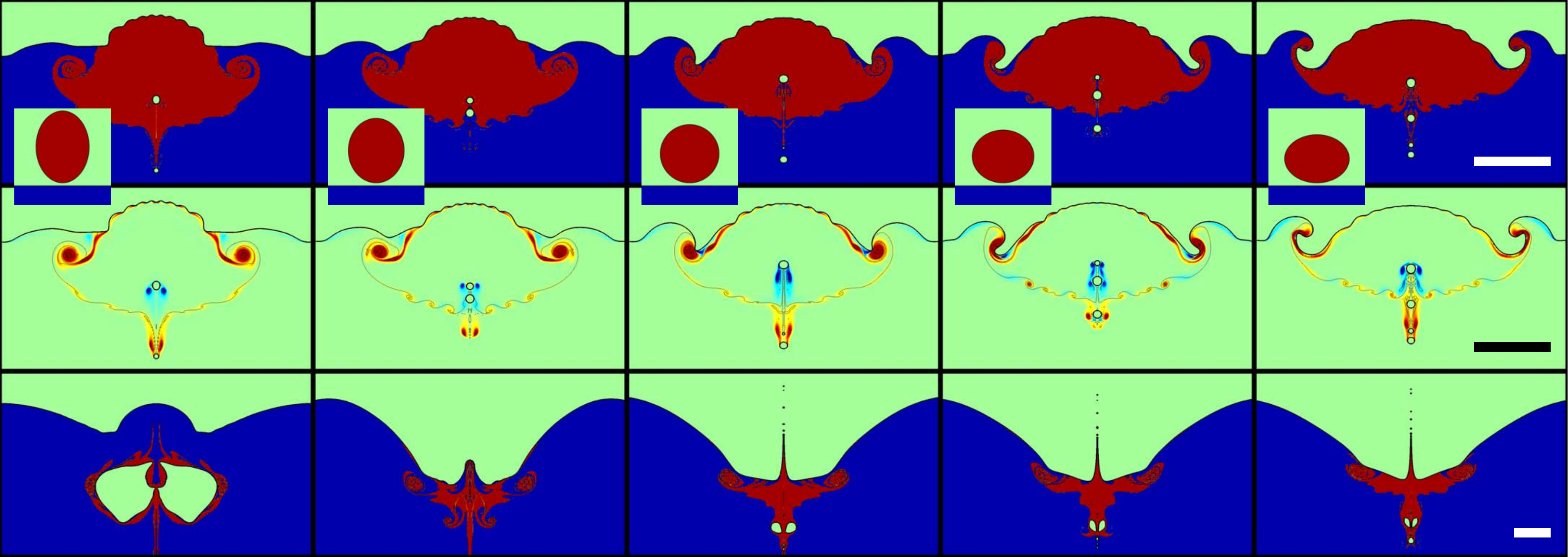}
  \caption{Effect of drop shape on the vortex interaction with the free surface: $D =$~5~mm, $V =$~1~m/s,
  over a range of aspect ratios from left to right $a =$~0.9, 0.95, 1, 1.05 and 1.1, which are shown in the insets (not to scale).
	First row: interface shapes at a fixed time $t/\tau_v = 0.72$.
	Middle row: vorticity in the liquid at the same time.
	Bottom row: interface shapes shown at a later time $t^{*} = 5.72$.
	The scale bars are 0.5~$D$ long.}
  \label{fig:compareTauv}
 \end{center}
\end{figure*}

To understand the mechanism responsible for the large bubble entrapment,
we focus from here on one typical impact condition ($D =$~5~mm, $V =$~1~m/s), and change the drop shape.
Figure~\ref{fig:compareTauv}, gives a clear indication of the role of the drop shape in the large bubble entrapment.
It changes the strength and dynamics of the vortex ring and its interaction with the interface.
The relevant time scale for the drop entry time into the pool and the early formation of the cavity
is $\tau_{v} = D_{v}/V = \tau / a^2$.
By comparing the impact for five different drop shapes at the same non-dimensional time $t/\tau_v$, we observe that
the vortex pulls on the interface and self-destructs earlier for the flatter drop (larger $a$).
It can therefore not pull on the interface to produce the large bubble.

\begin{figure}
 \begin{center}
  \includegraphics[width=\linewidth]{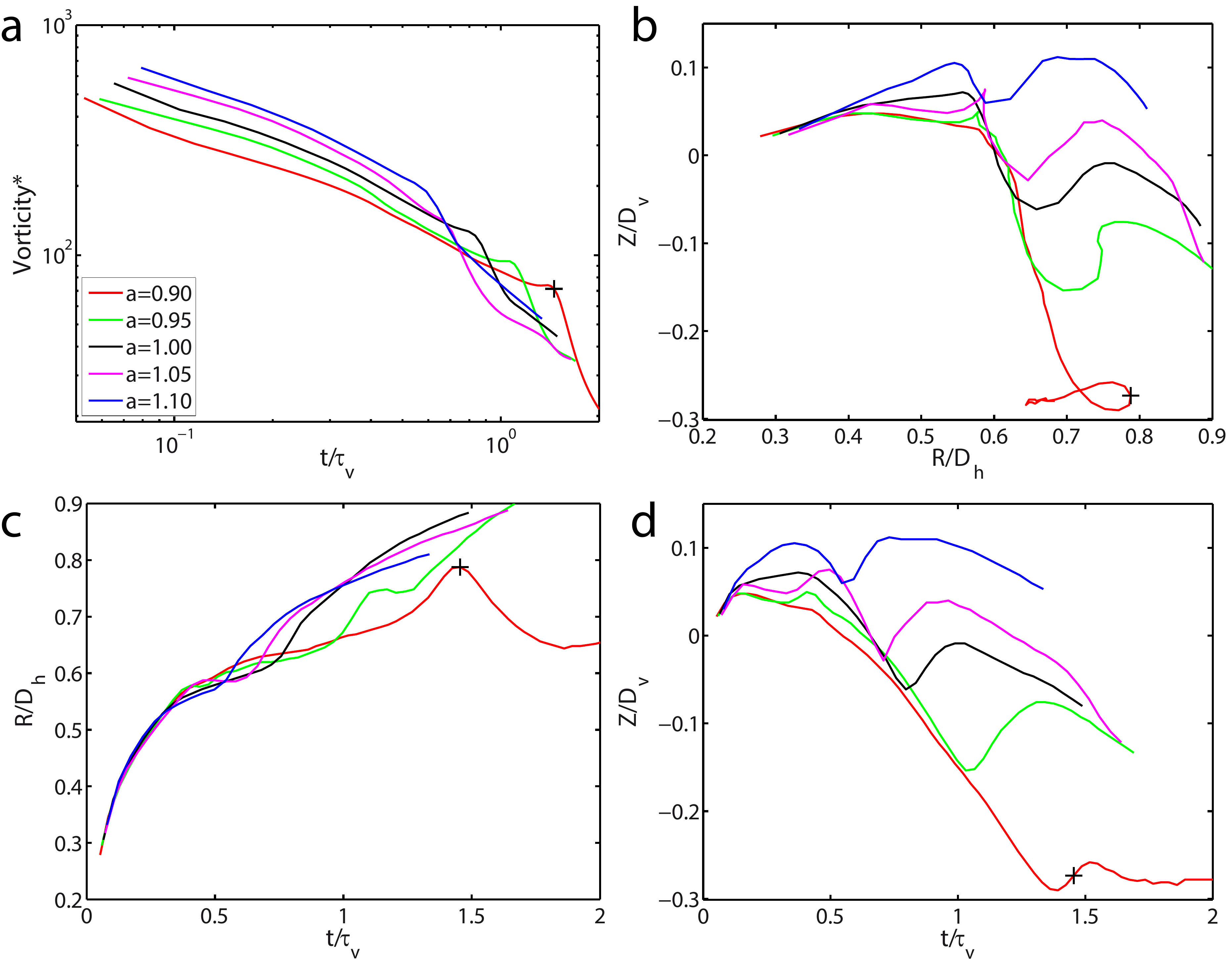}
  \caption{Time evolution of the vorticity maximum in the main vortex ring shed behind the neck (a),
	and its location (b), starting near the center and then moving outwards radially.
	(c) \& (d) show the time evolutions of the its radial and vertical location.
  The red curve represents the case where the large-bubble is entrapped.
	The \textbf{+} indicates the time of maximal radial location, corresponding to $t^{*} = $~1.80 (fifth panel in Fig.~\ref{fig:largeBubble}).}
  \label{fig:Vorticity}
 \end{center}
\end{figure}

This can be further analyzed by tracking the strongest vortex ring shed into the pool (Fig.~\ref{fig:Vorticity}),
as shown in Fig.~\ref{fig:largeBubble}(b) and the second row of Fig.~\ref{fig:compareTauv}.
The time when the vortex ring strongly pulls on the interface can be identified in its trajectory,
when it moves together with the deformed interface. This shows the vortex move rapidly radially outwards,
and then being deflected upwards, forming a characteristic hump in Figure~\ref{fig:Vorticity}(b).
As a consequence of this pulling, a tongue of the interface penetrates into the vortex ring,
leading to a sharp decrease of its intensity (Fig.~\ref{fig:Vorticity}a).
This also helps promote an inward motion, above the vortex, in the case of large bubble entrapment (starting from the \textbf{+} on the curve).

The surface pulling occurs later for more prolate drops, which confirms the previous observation from Fig.~\ref{fig:compareTauv}.
For $a = 0.9$, it even occurs for $t/\tau_V > 1$.
The location of the pulling is also very different.
For the flattest drop ($a = 1.1$) it occurs above the pool surface, while for $a = 0.9$ it is lower than $-D/3$.
This changes the curvature and orientation of the interface where the vortex pulls, leading to different crater structures.
For the oblate drop, the vortex starts to pull on the side of the drop before it has fully entered the pool.
The resulting liquid protrusion is thinner and more vertical, and therefore is rapidly pulled back by surface tension.
In contrast, the prolate drop pulls on the interface below the pool surface,
on the side of the cavity created by the impacting drop (fifth panel of Fig.~\ref{fig:largeBubble}).
The thicker and more horizontal liquid tongue is thus able to collapse on the central axis to entrap the large bubble.

\subsection{Vortex dynamics}

To understand how the shape of the drop changes so drastically the time and location of the pulling,
we need to analyze the earlier dynamics of the vortex ring.
First, we observe that a more oblate drop produces a stronger vortex ring (Fig.~\ref{fig:Vorticity}a).
This is due to a geometrical effect: when a flatter drop meets the pool surface, the neck shape forms a sharper cusp
than for a circular or prolate drop.
A stronger curvature will produce stronger vorticity (see section~\ref{sec:discussion}).
Then, we also observe that the dynamics of the vortex ring changes with the drop shape.
Although the vortex rings closely follows the neck in the radial direction in all cases
up to $t/\tau_v = 0.5$ (see Fig.~\ref{fig:VorticityRt} \cite{SM}), the dynamics is different in the vertical direction.
The vortex ring separates from the interface earlier for more prolate drops (Fig.~\ref{fig:Vorticity}d).

These two effects both contribute to the delayed pulling of the vortex ring on the interface for the prolate drop,
leading to the large bubble entrapment.
As the vortex is weaker and further away from the interface in that case, it exerts a smaller force on the interface.
It is therefore able to move deeper into the pool before pulling on the interface.

This explains why the pulling of the interface occurs later for prolate drops,
only when the opening cavity brings again the interface closer to the vortex.
It also explains why the vortex ring is able to generates a stronger vortical flow, as it is further away from the interface.
The key role of the vorticity shedding in the dynamics of drop impacts was first observed in Thoraval~\textit{et al.} \cite{Thoraval2012},
and further studied in \cite{Moore2014, Agbaglah2015, Lee2015}.
However, the complex process of vorticity production and separation from the free surface is still not fully understood.
This study brings another contribution to this problem, showing the effect of the drop shape on the vorticity shedding.

The earlier vorticity shedding for more prolate drops is probably due to
the faster vertical motion of the sharp interface point on the side of the drop
(wider angle between the drop and the pool).
As the liquid from the pool moves faster vertically on the side of the drop (see Fig.~\ref{fig:largeBubble}),
the corner where vorticity is generated is pushed against the dominant downwards velocity, thus promoting vorticity shedding.

\subsection{Energy and momentum}

\begin{figure}
 \begin{center}
 \includegraphics[width=0.8\linewidth]{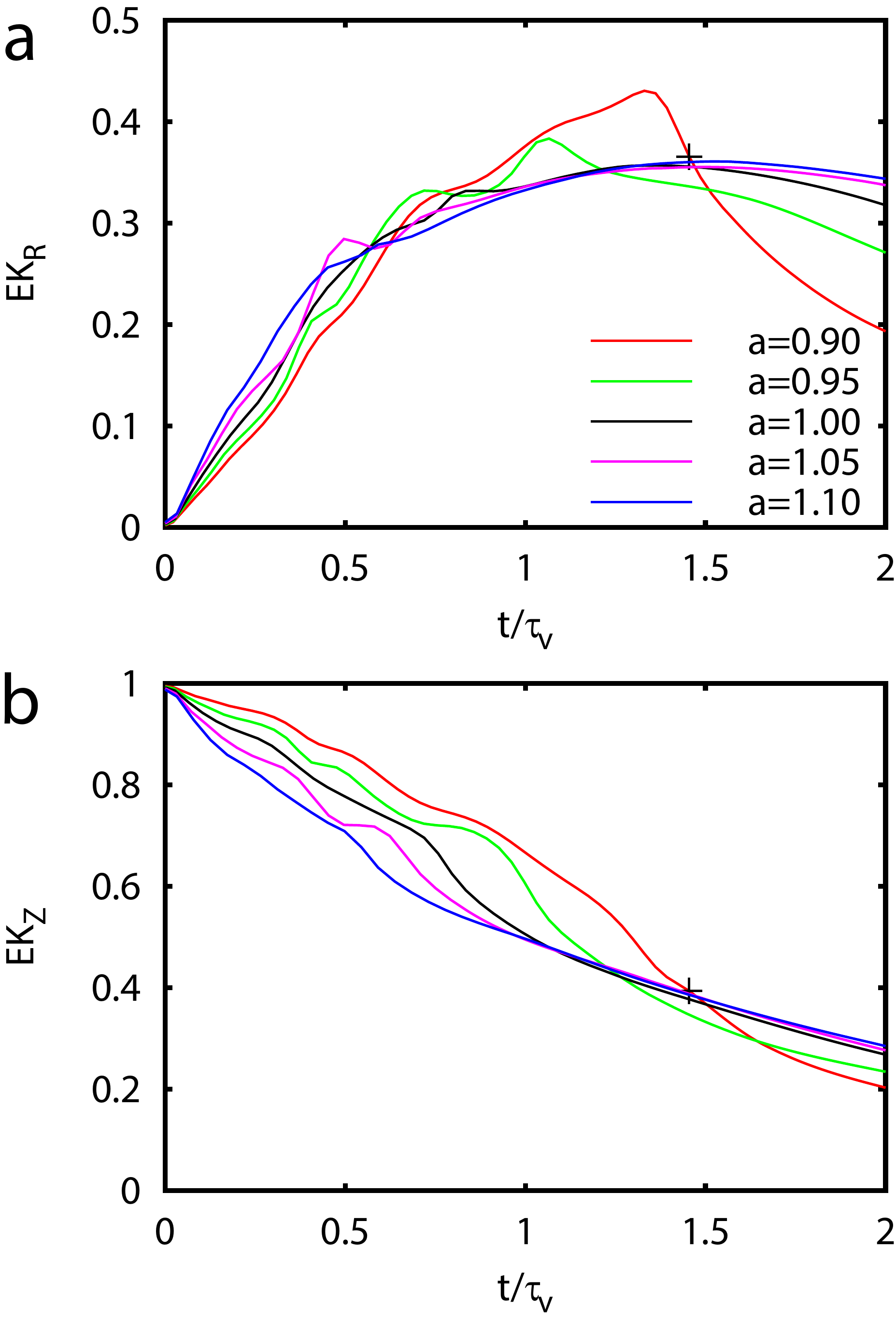}
  \caption{Kinetic energy $EK$ in radial (a) and vertical (b) directions,
	normalized by the initial kinetic energy of the drop.
The total combined kinetic energy is shown in Supplementary Figure~\ref{fig:EK} \cite{SM}.}
  \label{fig:Ek}
 \end{center}
\end{figure}

To understand the effect of the vortex on the larger-scale dynamics,
we can decompose the kinetic energy and momentum in the radial and vertical directions (Fig.~\ref{fig:Ek}).
The radial (resp. vertical) kinetic energy $EK_R$ (resp. $EK_Z$) is defined as the volume integral over the whole domain of
$\frac{1}{2} \rho V_R^2$ (resp. $\frac{1}{2} \rho V_Z^2$), where $V_R$ (resp. $V_Z$) is the radial (resp. vertical) component of the velocity.
The total kinetic energy is the sum of the two for this axisymmetric simulation: $EK = EK_R + EK_Z$.

The dominant energy in the impact is the kinetic energy, while the initial surface and potential energy
represent only 17\% and 10\% respectively for the typical case studied here.
The surface energy of the ellipsoidal shapes is larger than the one of the spherical drop, as it is the minimal surface.
However, for the shapes studied here, the surface energy changes by less than 1.7\%,
and therefore less than 0.3\% of the initial kinetic energy.
This difference can therefore be neglected.
The kinetic energy is initially only vertical, and the radial kinetic energy is produced as the drop enters the pool.

We also normalize the time by $\tau_{v}$ based on the vertical diameter,
as it is the relevant time scale at which the energy is transferred from the impacting drop to the pool.
Even in this rescaled time frame, the release of vertical kinetic energy
is initially slower for the drops with a larger vertical extension.
This can be understood by an added mass effect, as the resistance from the pool will be lower
for the drops showing a smaller horizontal extent perpendicular to the impact direction.
However, this tendency is later reversed, so that all cases reach a similar
vertical kinetic energy around $t/\tau_{v} \geq 1.5$.

\begin{figure}
\begin{center}
	\includegraphics[width=0.8\linewidth]{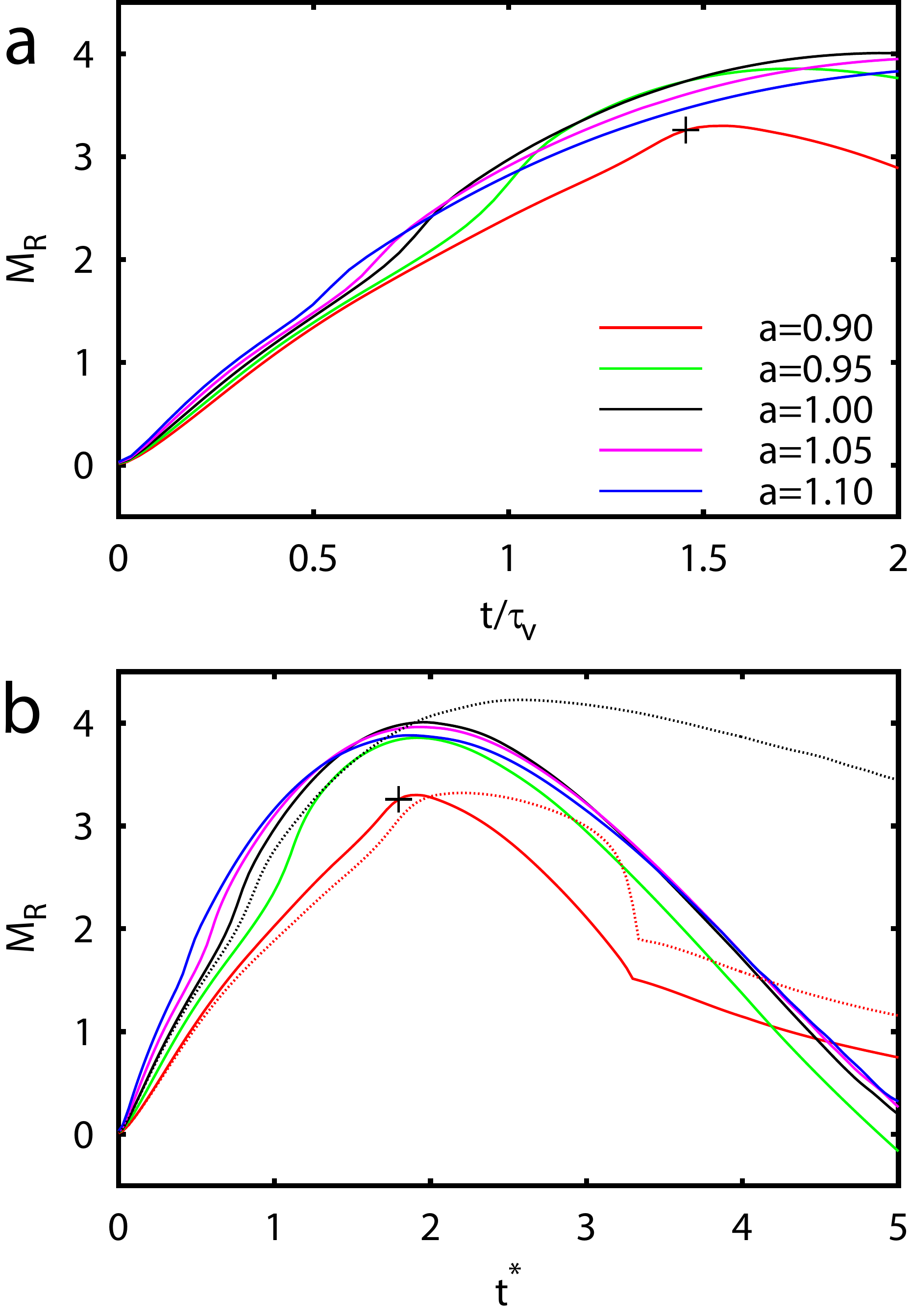}
	\caption{Radial component of the momentum $M_R$,
	normalized by the norm of the initial vertical momentum of the drop.
	The dotted lines in (b) correspond to the cases where gravity has been removed.}
	\label{fig:MR}
\end{center}
\end{figure}

From $t/\tau_{v} \geq 0.75$ until the pulling time of the vortex on the interface,
the radial kinetic energy is larger for the prolate drop (Fig.~\ref{fig:Ek}a).
However, the outwards radial momentum is still lower (Fig.~\ref{fig:MR}a).
The larger kinetic energy therefore reveals a stronger inward velocity towards the axis of symmetry,
as the dominant radial flow is away from the center.
This shows that although the vortex for the prolate drop is initially weaker,
it is able to generate a stronger large-scale rotational motion in the flow,
beneficial for the closure of the liquid tongue above the large bubble.
In contrast, the stronger vortex for the oblate drop self-destructs earlier by pulling on the interface,
stopping the development of the vortical flow.

Finally, Fig.~\ref{fig:MR}(b) shows that the radial momentum starts decreasing strongly after $t^{*} \simeq 2$,
non-dimensionalized by $\tau = D/V$, based on the equivalent diameter of the drop.
This is characteristic of the reversing flow due to gravity towards the axis of symmetry,
as confirmed by the comparison with the same cases without gravity.
Gravity starts playing a significant role from that time, as shown by the large deviation of the dotted curves.

This reversing flow is probably beneficial for the liquid tongue to close the cavity and entrap the large bubble.
As the interface pulling occurs later for the prolate drop (also in the rescaled time with $\tau_{v} = D_v/V$),
it is closer to the time when the liquid flows back toward the axis.
However, even without the effect of gravity, the radial momentum still shows a sharp decrease
just before the large bubble entrapment.
It therefore shows that the closure of the liquid tongue above the cavity
is mostly driven by surface tension, after the strong decrease of radial expansion by the vortical flow.

\subsection{Other experimental conditions}
\label{sec:Other}

\begin{figure}[h!]
 \begin{center}
  \includegraphics[width=1.0\linewidth]{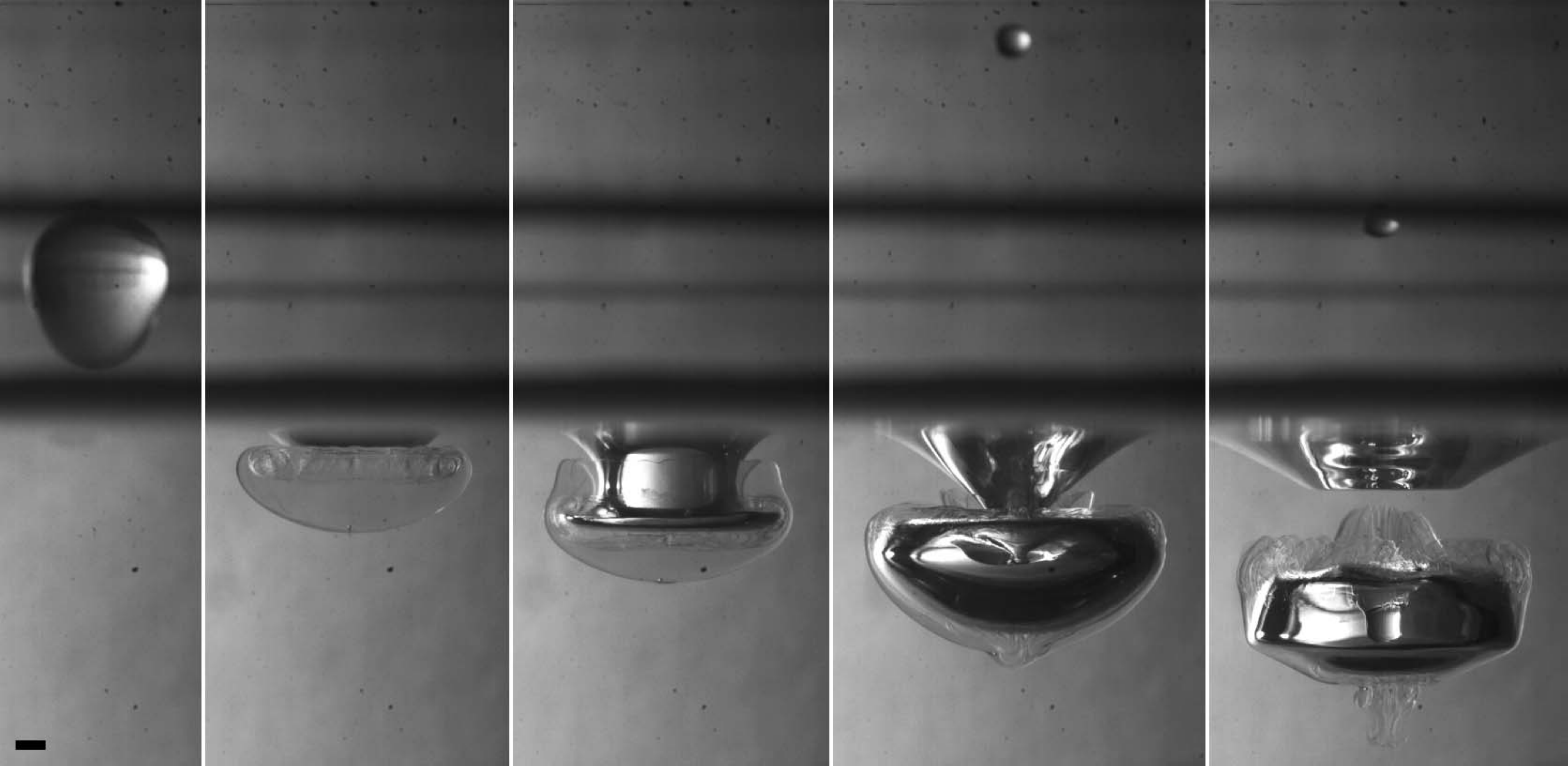}
  \caption{Large bubble entrapment for a drop of 6\% $MgSO_4$, giving $\rho_d / \rho_p =$~1.06 and $\mu_d =$~1.3~cP.
	for $D=$5.0 mm and $V=$~0.94~m/s, $H=$~5.0~cm.
  The sequence shows the following times: -2.3, 7, 11, 17, 25~ms after the first contact.
	The scale bar is 1~mm long.}
  \label{Fig_S1}
 \end{center}
\end{figure}

Figure~\ref{fig:LBexperiments} uses 10.5\% $MgSO_4$ in the water drop
to introduce a strong index-of-refraction difference between the drop and the pool liquids,
to allow us to visualize clearly their interface and identify vortex structures.
This also introduces a density difference $\rho_d / \rho_p =$ 1.105 and viscosity difference,
where the drop liquid is 1.95 times more viscous than the distilled water in the pool.
However, in the numerical simulations, both liquids are modeled with water properties.
To quantify possible effects of the difference in properties,
we performed another set of experiments with about half the salt concentration,
i.e. for 6\% $MgSO_4$.  Figure~\ref{Fig_S1} shows that the overall entrapment of a large bubble
is the same as for the higher salt concentration.
This suggests that the density difference between the drop and pool does not significantly change the
large-bubble entrapment dynamics.
The lower concentration reduces the contrast in the imaging.

\begin{figure}
 \begin{center}
	\includegraphics[width=1.0\linewidth]{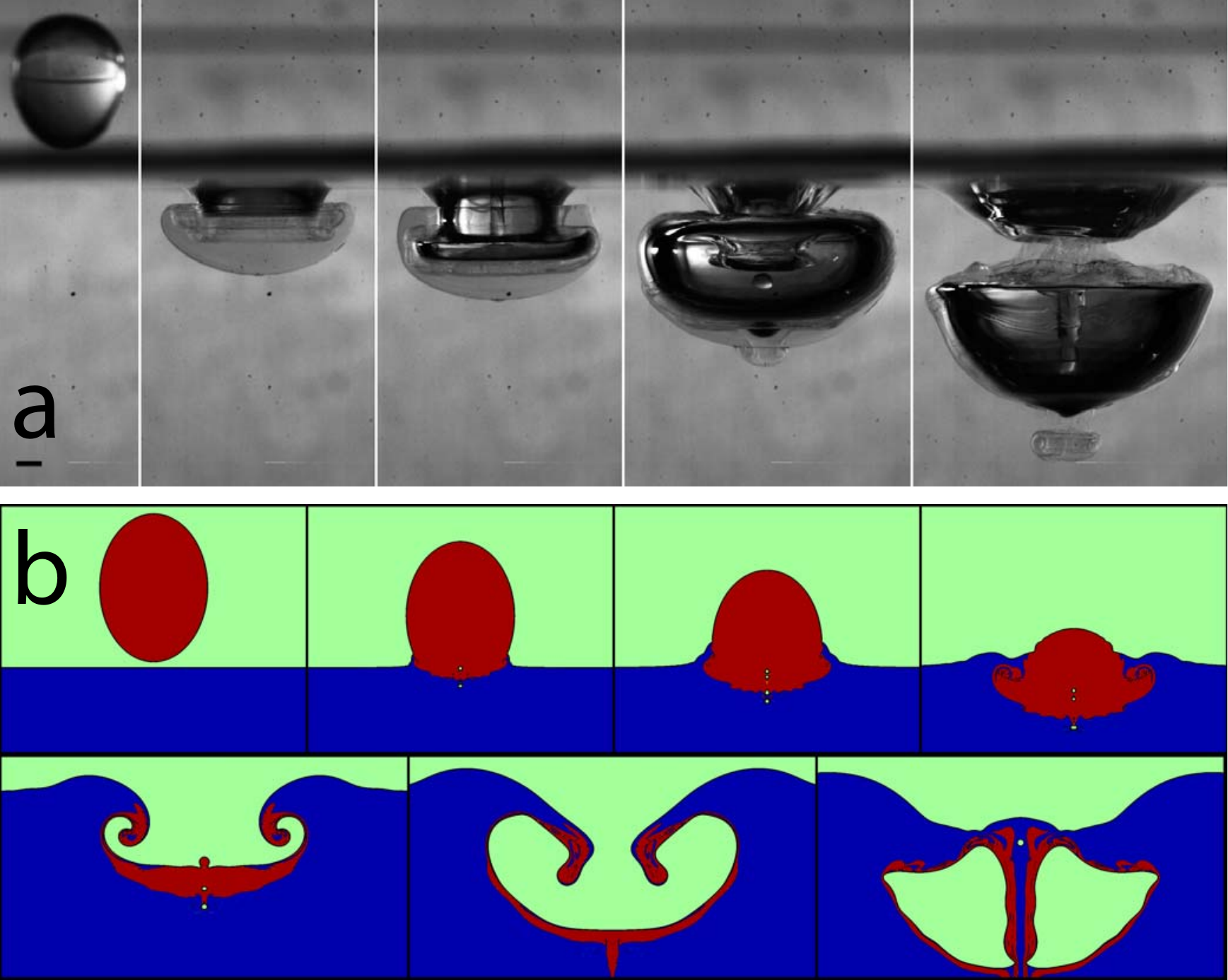}
  \caption{(a) Large bubble entrapment for a drop of 10.5\% $MgSO_{4}$ solution with viscosity 1.96~cP and density 1.105~g/cm$^3$,
	impacting into a pool of distilled water.
	Impact conditions: Prolate drop, $D=$~5.0~mm, $V=$~1.30~m/s, $H=$~9.1~cm.
  The sequence shows the following times: -1.3, 5.4, 7, 11, 20~ms after the first contact.
	The scale bar is 1~mm long.
	(b) Sequence in the same conditions as in Fig.~\ref{fig:largeBubble},
	but higher impact velocity, and thus similar conditions as in the experiments of Fig.~\ref{Fig_S2}.
	$t^{*} = $~-0.05, 0.18, 0.41, 0.87, 1.80, 2.99, 4.80.
	$D =$~5~mm, $V =$~1.30~m/s, $a =$~0.9.}
  \label{Fig_S2}
 \end{center}
\end{figure}

Figures~\ref{Fig_S2}(a) shows the entrapment dynamics for a prolate drop with the same liquid properties as Fig.~\ref{fig:LBexperiments},
but at a larger impact velocity.
The numerical simulation for similar conditions again reproduces very well the dynamics,
including the shape of the large bubble and the liquid column at its center (Fig.~\ref{Fig_S2}b).

\begin{figure}
 \begin{center}
	\includegraphics[width=1.0\linewidth]{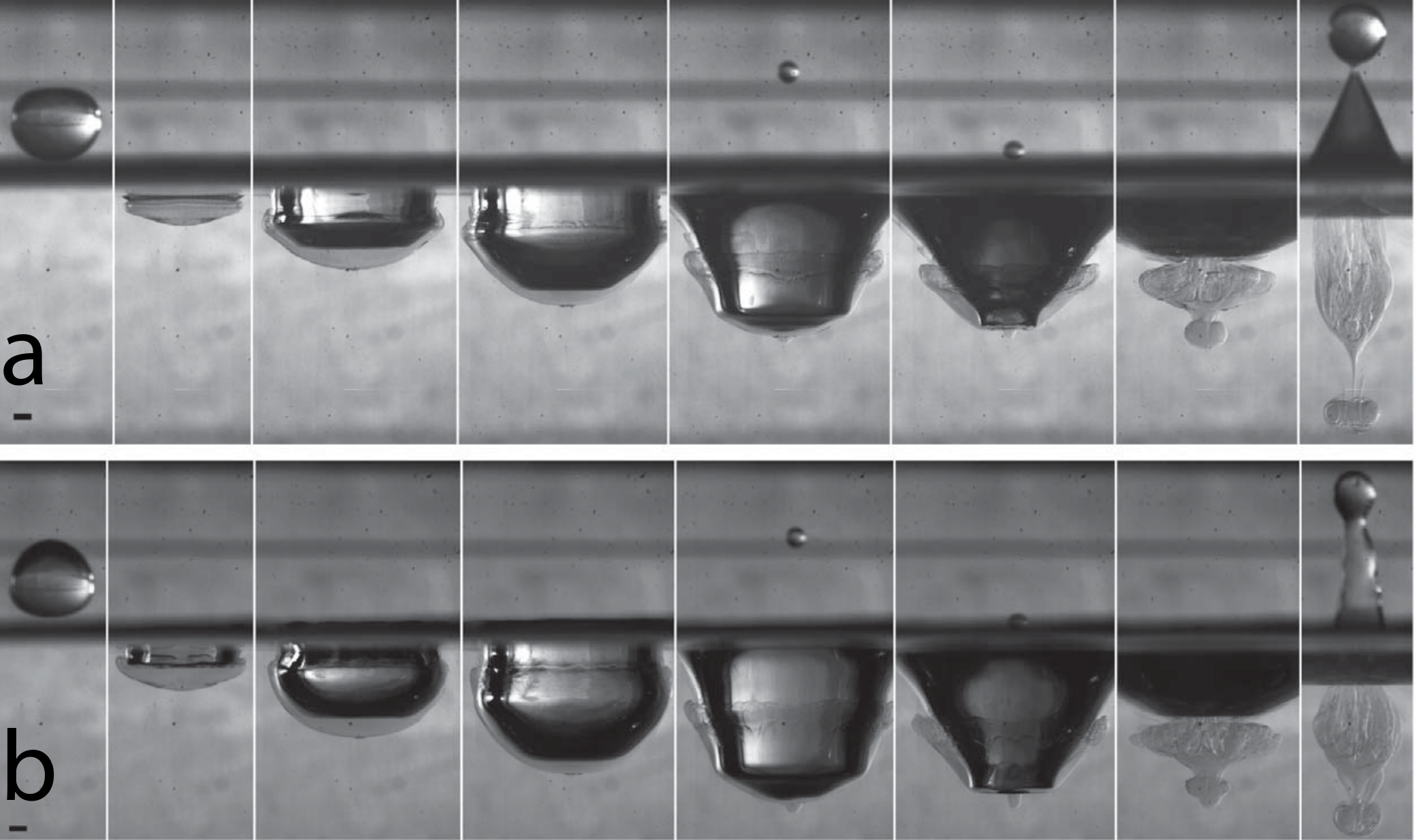}
  \caption{Crater evolution for oblate drops, which never entrap a large bubble.
	Drop composition: 10.5\% $MgSO_{4}$ solution with viscosity 1.96~cP and density 1.105~g/cm$^3$; pool: distilled water.
	Impact conditions: Oblate drop, $D=$~5.0~mm.
  (a) For $V=$~1.11~m/s, $H=$~6.8~cm.  The sequence shows the following times: -2, 2.7, 7.7, 12.7, 24.3, 29, 37, 67.7~ms after the first contact.
  (b) For $V=$~1.24~m/s, $H=$~8.3~cm.  The sequence shows the following times: -1.5, 3.5, 8.5, 13.5, 25.2, 29.8, 37.8, 50.8~ms after the first contact.
	The scale bars are 1~mm long.}
  \label{Fig_S3}
 \end{center}
\end{figure}

For reference, experiments in Figure~\ref{Fig_S3} show the dynamics for oblate drops at two intermediate impact velocities,
where no large bubble is entrapped.
However, the collapse of the crater can produce vortex rings, as visible in the last panels.
The small drop visible in the fifth panel of both sequences is the satellite
produced when the primary drop pinches off from the nozzle.

\section{Discussion}
\label{sec:discussion}

Herein we have argued that the vortex ring produced by the early contact of the drop with the pool
can explain when a large-bubble is entrapped into the pool.
It is therefore crucial to explain how this vortex is formed and how its formation relates to the drop shape.
This issue of vorticity production was debated two decades ago in the context of drop impact
\cite{Morton1992, Cresswell1992, Peck1994, Cresswell1995, Dooley1997},
but appears not to have been universally accepted within the community.
The proposed mechanism was neither verified experimentally nor numerically,
due to the challenges of capturing the dynamics near the interface.
It is therefore worthwhile to review some of the underlying arguments behind the theory,
and support it quantitatively with our numerical results.
A general review on vorticity production can be found in Brøns \textit{et al.} \cite{Brons2014},
while we will focus here only on the case of drop impact.

\begin{figure*}
 \begin{center}
  \includegraphics[width=0.8\linewidth]{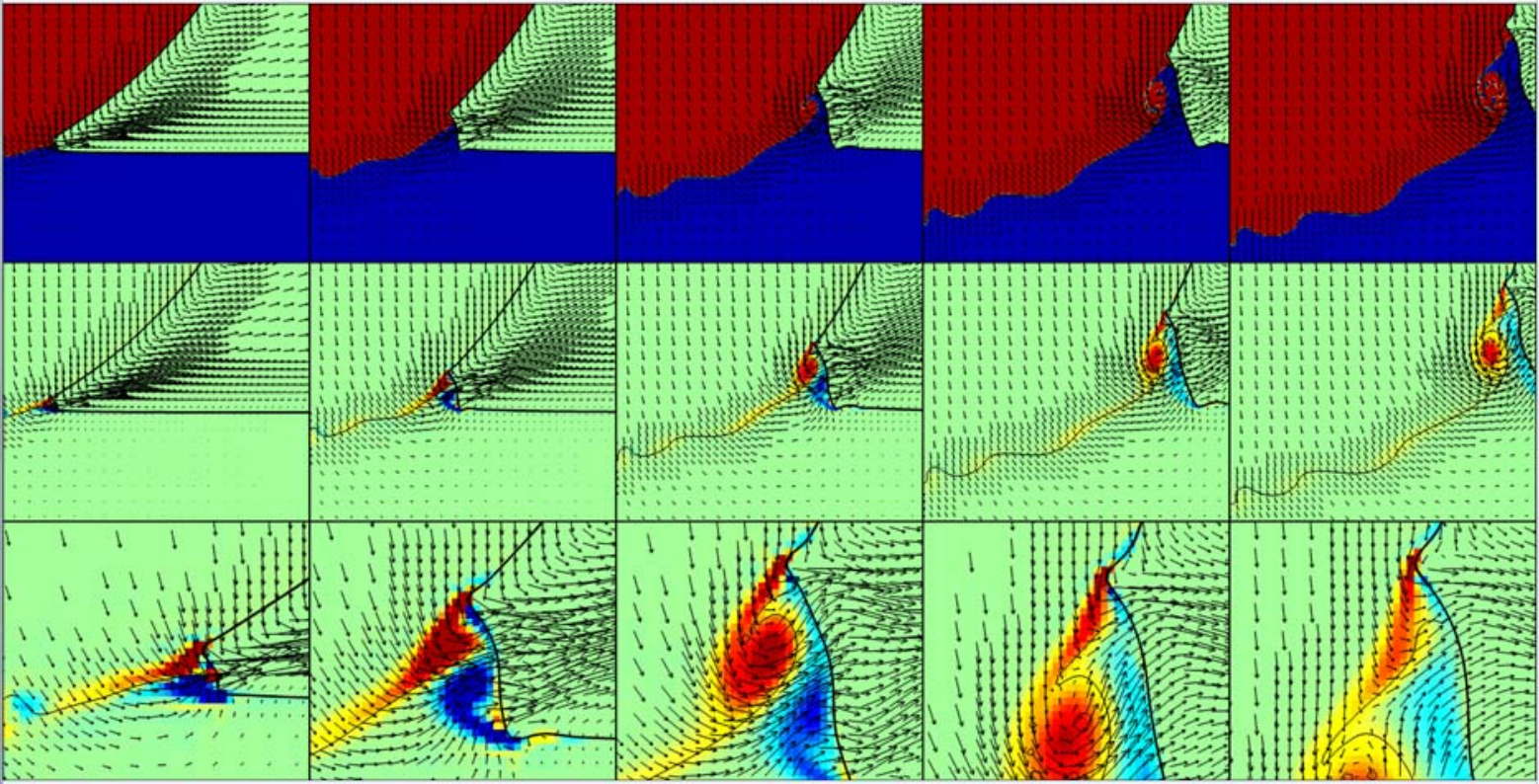}
  \caption{Emergence of the vorticity from the growing neck connecting the drop and the pool.}
  \label{Fig_15}
 \end{center}
\end{figure*}

\subsection{Vorticity production in drop impact}

The basic concept of {\it the persistence of irrotationality}, i.e. fluid originally in irrotational motion
will continue to move irrotationally, is one of the foundations of classical hydrodynamics.
This theorem is discussed by Batchelor (page 276-277 in \cite{Batchelor1967}) and he attributes it to Lagrange,
with the first demonstration by Cauchy in 1815.
The theorem applies for homogeneous flows with constant density which are originally irrotational or start from rest.
This applies at all times for inviscid fluids, but also for regular viscous fluids away from the boundaries,
from where the vorticity must emerge.
Batchelor further states: {\it ''Vortex strength or circulation cannot be created in the interior of the fluid,
but once there it is spread by the action of viscosity''}
and he continues for motions of a viscous fluid from rest (page 277 in \cite{Batchelor1967}):
{\it ''Initially the vorticity is everywhere zero and the motion must remain wholly irrotational
unless vorticity diffuses across the surface bounding the fluid''}.

This is clearly the basic condition when a drop impacts on a pool.
The pool is at rest and the motions within the falling drop are slow and likely well described by basic irrotational shape-oscillations,
which are promoted during the pinch-off from the nozzle.
In the numerical simulations these initial motions within the drop can be set to zero and our initial conditions are simply a drop with a uniform vertical velocity.
With this initial condition, on the drop flow-field, our numerical simulations reproduce the experimental results.
We therefore conclude that the oscillatory motions within the falling drop are not a significant factor during the very rapid impact motions and can be ignored.
The oscillations only served the purpose of changing the drop shapes present at initial contact.

The vorticity in the observed vortex ring must therefore arise from the free surface and
cannot be produced spontaneously within the pool by the pressure-driven initial flow due to the impact.
Those motions must be irrotational as postulated above.
This can also be inferred from the absence of the pressure in the vorticity equation.

Next we must ask: what is the nature of the vorticity production?

Cresswell \& Morton \cite{Cresswell1995} propose what we believe is the correct origin of the vorticity.
Their explanation is based on the vorticity produced by the flow $q$ along a curved free surface,
which has a local radius of curvature $R$.
The strength of this vorticity generation for a stationary two-dimensional free surface has the form
(Batchelor, page 366 \cite{Batchelor1967}, \cite{Sebilleau2009}):
\begin{equation}
\Delta \omega = \frac{2q }{R}
\label{Eq_1}
\end{equation}
This equation essentially arises from the fact, that we cannot satisfy simultaneously the stress-free boundary condition and irrotationality
and must relax the latter to avoid infinite tangential accelerations of the fluid element at the free surface.

The free surface can be considered circular locally by expansion of the Navier-Stokes equations \cite{Lugt1987}.
In the corresponding local polar coordinates, the two conditions are expressed by the sum and difference of identical terms,
\[
\sigma_{r\theta} = \frac{\partial u_{\theta}}{\partial r} - \left( \frac{u_{\theta}}{r} - \frac{1}{r} \frac{\partial u_r}{\partial \theta} \right) = 0
\]
\[
\omega = \frac{\partial u_{\theta}}{\partial r} + \left( \frac{u_{\theta}}{r} - \frac{1}{r} \frac{\partial u_r}{\partial \theta} \right) = 0\;\; 
\]
which cannot both be satisfied at the same time.
Substituting into the $\omega$ equation for $\partial u_{\theta} / \partial r$ from the first equation
and realizing that $\partial u_r / \partial \theta=0$ along the circular free surface,
we get $\omega = 2u_r/r$, thereby recovering Eq.~\ref{Eq_1}.
For strong vorticity production, either $q$ must be large, or the radius of curvature $R$ very small.
For the drop-impact case, the sharp corner is formed in the neck connecting the drop to the pool
once they have formed a bridge.

Eq.~\ref{Eq_1} is also valid for an interface with a fixed shape moving at a constant velocity \cite[p. 365]{Batchelor1967}.
Ohring and Lugt \cite{Ohring1989, Ohring1991} have also shown that it also gives a correct estimate
of the vorticity produced for slowly deforming interfaces.
Equation~\ref{Eq_1} has been generalized to flows along three-dimensional free surfaces by Peck \& Sigurdson \cite{Peck1998b},
but this is not needed here.

Although Eq.~\ref{Eq_1} explains the origin of the vorticity, it only appears as a boundary condition to satisfy on the interface.
Once this vorticity is produced, it will diffuse into a thin boundary layer, which can separate to enter the pool.
These mechanisms have been studied by \cite{Dooley1997, Peck1999}.
However, it still remains a challenge to understand the separation of vorticity in drop impacts
and its effects on the interface \cite{Moore2014, Agbaglah2015}.

\subsection{Validation}

Our numerical simulations have been able to correctly reproduce the large bubble entrapment
that was previously observed experimentally, and explain it by the pulling of a vortex ring.
They can therefore be used to pinpoint the origin of the vorticity in this vortex ring.

The insets in Fig.~\ref{fig:largeBubble} show that the vortex ring emerge from the sharp upper corner in the rapidly growing neck.
This is consistent with the explanation of Cresswell \& Morton \cite{Cresswell1995},
suggesting that the largest vorticity should be created in the regions of highest curvature.
This is shown in more detail in Fig.~\ref{Fig_15}.
Near the center of the drop the velocity vectors are vertical and gradually turn towards horizontal.

We can make another important clarification with this figure:
no vorticity is produced at the interface between the liquid originally in the drop or the pool.
In the bottom left panel, we observe that two vortices of opposite signs are produced in the upper and lower corner of the neck.
This is consistent with the previous results of Josserand \& Zaleski \cite{Josserand2003}.
In the last bottom panel, the top corner of the neck has moved further away from the line,
together with the point of highest vorticity, demonstrating this even more clearly.
For much higher impact velocities than studied here, the two opposite sign vortices
can shed alternatively forming a vortex street \cite{Thoraval2012}.

However, no significant vorticity is observed near the line separating the drop and pool liquids.
This simply comes from the fact that this interface, between the red and blue regions in top of Figure~\ref{fig:largeBubble}(a),
is not a material interface.
It is only a visualization artifact to identify the origin of the liquid, from the drop or pool.
In reality, it is only a homogeneous liquid.
The physical reason behind is that we are solving the Navier-Stokes equations,
which do not allow for discontinuity.
The contact between the drop and the pool occurs smoothly, due to the air cushioning.
The velocity field is continuous even across this line between drop and pool liquids,
without any vortex sheet produced there.

We can also verify the proposed vorticity production mechanism quantitatively with our numerical simulations.
Relation~\ref{Eq_1} is derived for a two-dimensional stationary free surface and the vorticity production could have other contributions
from the deforming or accelerating interface, or due to the axisymmetric geometry or even three-dimensional effects in the experiments,
or due to the presence of air.
However, those contributions will only change the strength of the vorticity but not where it originates, i.e. at the interface.
This can be verified in the numerical simulations, by estimating the terms in Eq.~\ref{Eq_1}.

\begin{figure}
 \begin{center}
  \includegraphics[width=\linewidth]{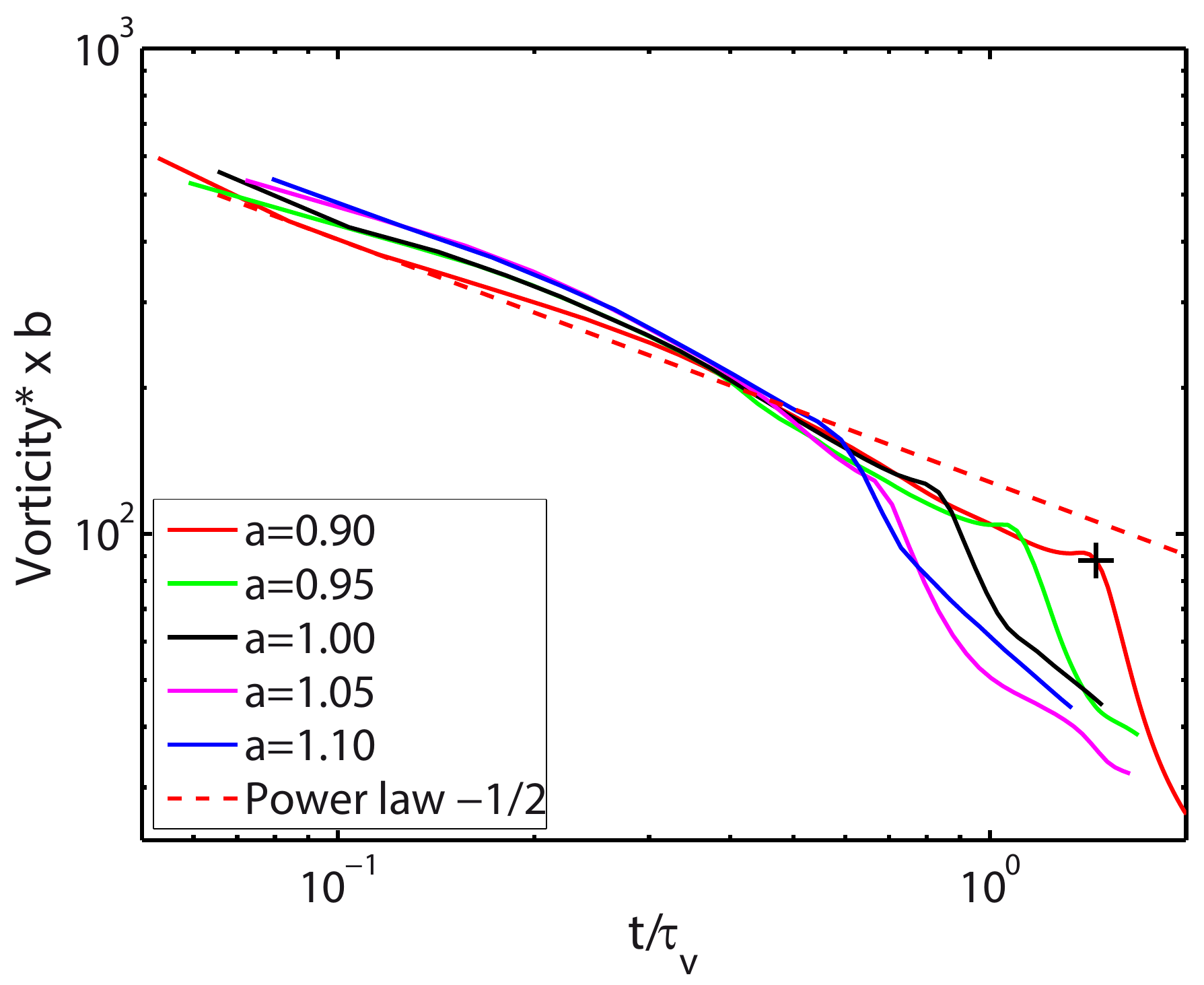}
  \caption{Strength of the vortex ring as in Fig.~\ref{fig:Vorticity}(a)
	but rescaled with the vertical extension of the drop $b$.}
  \label{fig:Vorticity-xb}
 \end{center}
\end{figure}

For a rough estimate we take the typical example of a water drop of diameter $D=5$~mm impacting at a velocity $V = 1$~m/s.
The vorticity maximum near the interface is $1.4 \times 10^5$~s$^-1$ at $t = 0.52$~ms.
The minimum radius of curvature of the interface is observed at the same point, with $R = 10$~$\mu$m,
and the tangential velocity relative to the interface is $q = 1.27$~m/s.
The right hand side of Eq.~\ref{Eq_1} $2q / R = 2.5 \times 10^5$
is of the same order of magnitude as the observed vorticity at the interface.

Another simple verification of Eq.~\ref{Eq_1} can be done by looking at the effect of the geometry.
Rescaling the strength of the vortex ring by the vertical extension of the drop $b$
collapses the curves as can be seen in Fig.~\ref{fig:Vorticity-xb}, 
where the curves are identical up to the point where the vortex interacts strongly with the free surface.
This is consistent with Eq.~\ref{Eq_1} where the vorticity is proportional to the curvature of the interface, $\propto 1/b$.
The early decay of the vorticity follows a $t^{-1/2}$ power-law, characteristic of viscous spreading.

\section{Conclusions}

Combining experimental and numerical approaches, we have explained the mechanism underlying the large-bubble entrapment during drop impact,
by a vortex ring which pulls the interface below the pool surface.
This pulling occurs earlier for oblate drops, leading to the self-destruction of the vortex.
In contrast, prolate drops produce a weaker vortex, that can separate from the interface during the early dynamics,
and develop a vortical flow below the pool surface.
As the drop enters the pool and produces a cavity, the interface is pulled by the vortex,
leading to the entrainment of the large bubble.

By turning off gravity in the numerical simulations,
we also show that hydrostatic pressure plays no role in the large-bubble entrapment.
Such bubble entrapment is observed for large drops impacting at low velocities,
relevant for the impact of secondary drops breaking from Worthington jets \cite{Rein1993, Liow2001, Fedorchenko2004a},
and wave breaking phenomena \cite{Andreas1995, Anguelova1999, Wanninkhof2009, Bird2010, Kiger2012, Lhuissier2012, Veron2012, Veron2015}.
It has therefore important environmental implications \cite{Bourouiba2013, Wilson2015},
but also in the industry where it can be beneficial for liquid aeration and gas-liquid chemical reactors \cite{Oguz1995},
and to enhance nucleate boiling \cite{Dhir1998}.

\bibliography{References}

\onecolumngrid
\appendix

\clearpage
\newpage

\setcounter{equation}{0}
\setcounter{figure}{0}
\setcounter{table}{0}
\setcounter{page}{1}
\makeatletter
\renewcommand{\theequation}{S\arabic{equation}}
\renewcommand{\thefigure}{S\arabic{figure}}
\renewcommand{\thetable}{S\arabic{table}}
\renewcommand{\thepage}{S\arabic{page}}

\onecolumngrid
\section*{Supplemental Material}

\subsection{Effect of computational refinement}

\begin{figure}[h!]
 \begin{center}
	\includegraphics[width=0.8\linewidth]{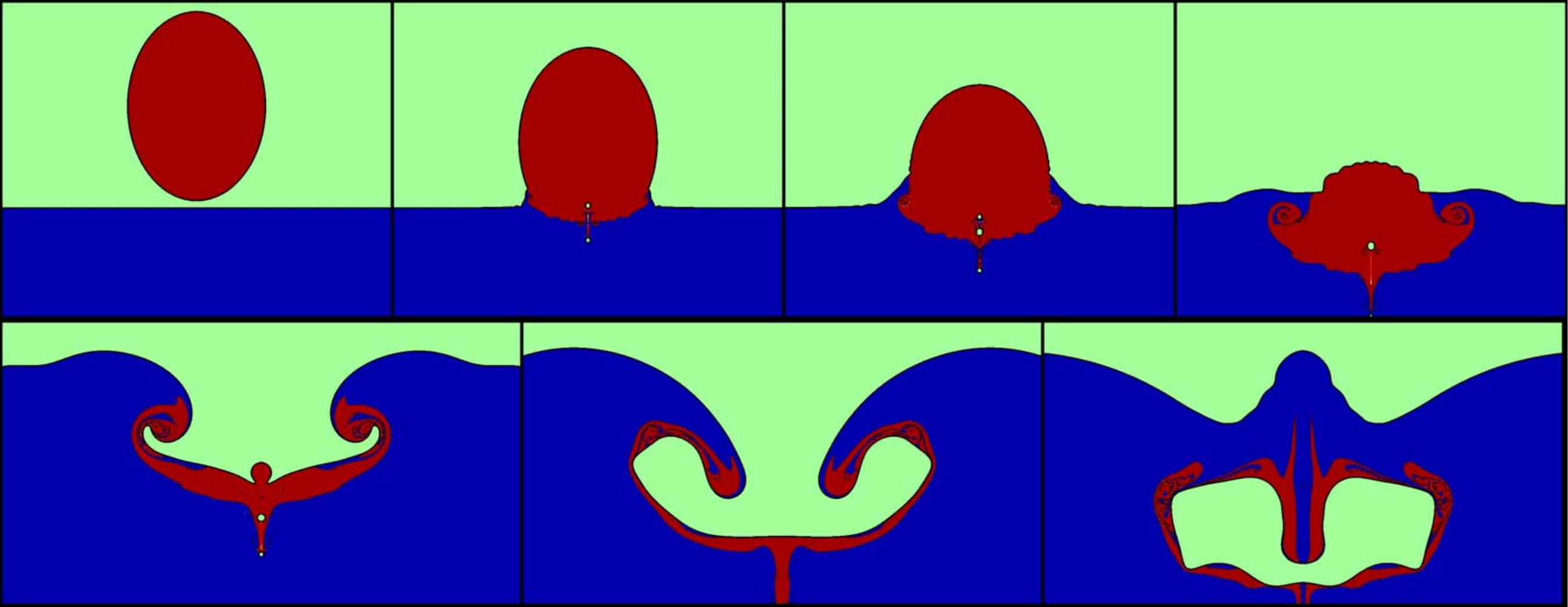}
  \includegraphics[width=0.8\linewidth]{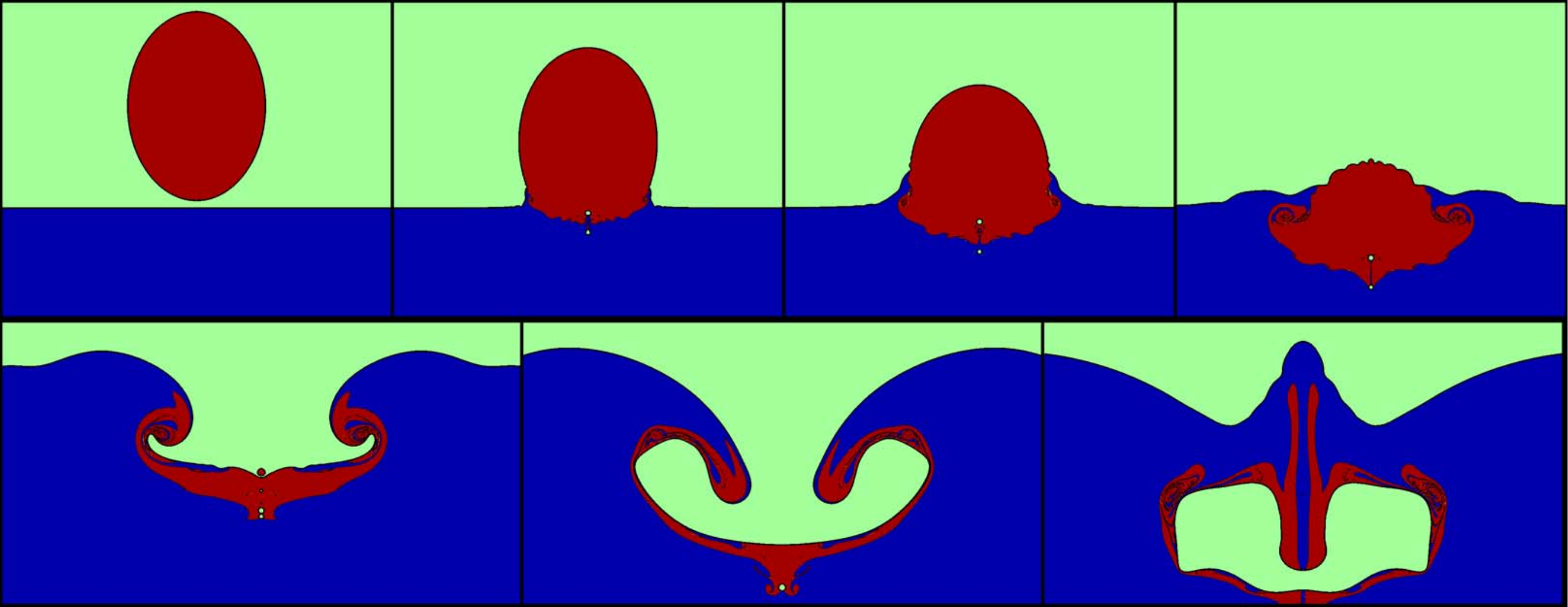}
  \caption{Sequence of simulations in the same conditions as in Fig.~\ref{fig:largeBubble},
	with maximum level of refinement 11 (top: $ D / \delta x = 532$) or 12 (bottom: $ D / \delta x = 1065$).
	$t^{*} = $~-0.05, 0.18, 0.41, 0.87, 1.80, 2.99, 4.80.
	$D =$~5~mm, $V =$~1.0~m/s, $a =$~0.9.}
  \label{fig:largeBubble-lmax12}
 \end{center}
\end{figure}

The large refinement of the grid chosen here was necessary to capture the large bubble entrapment,
because the vorticity production occurs in a small region near the pool surface.
In the case presented in Fig.~\ref{fig:largeBubble}, a higher level of refinement did not change the dynamics (see Fig.~\ref{fig:largeBubble-lmax12}).

\begin{figure}
 \begin{center}
	\includegraphics[width=0.8\linewidth]{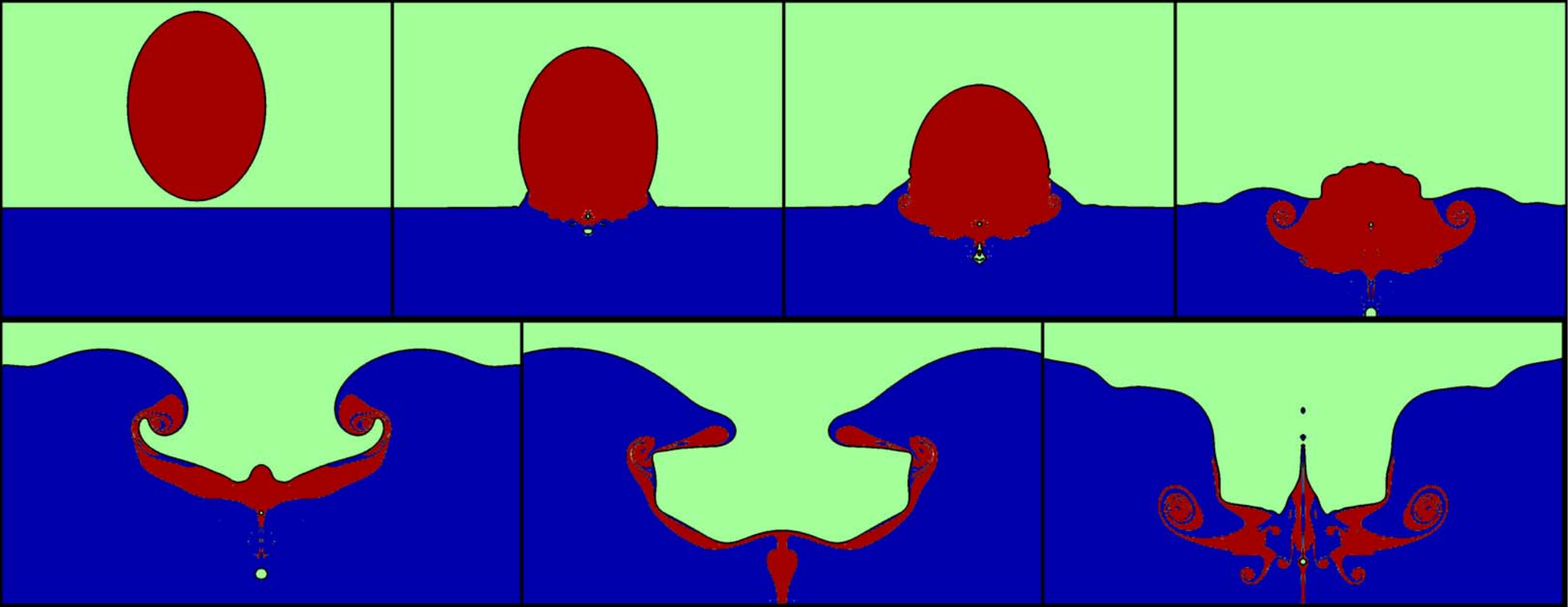}
  \includegraphics[width=0.8\linewidth]{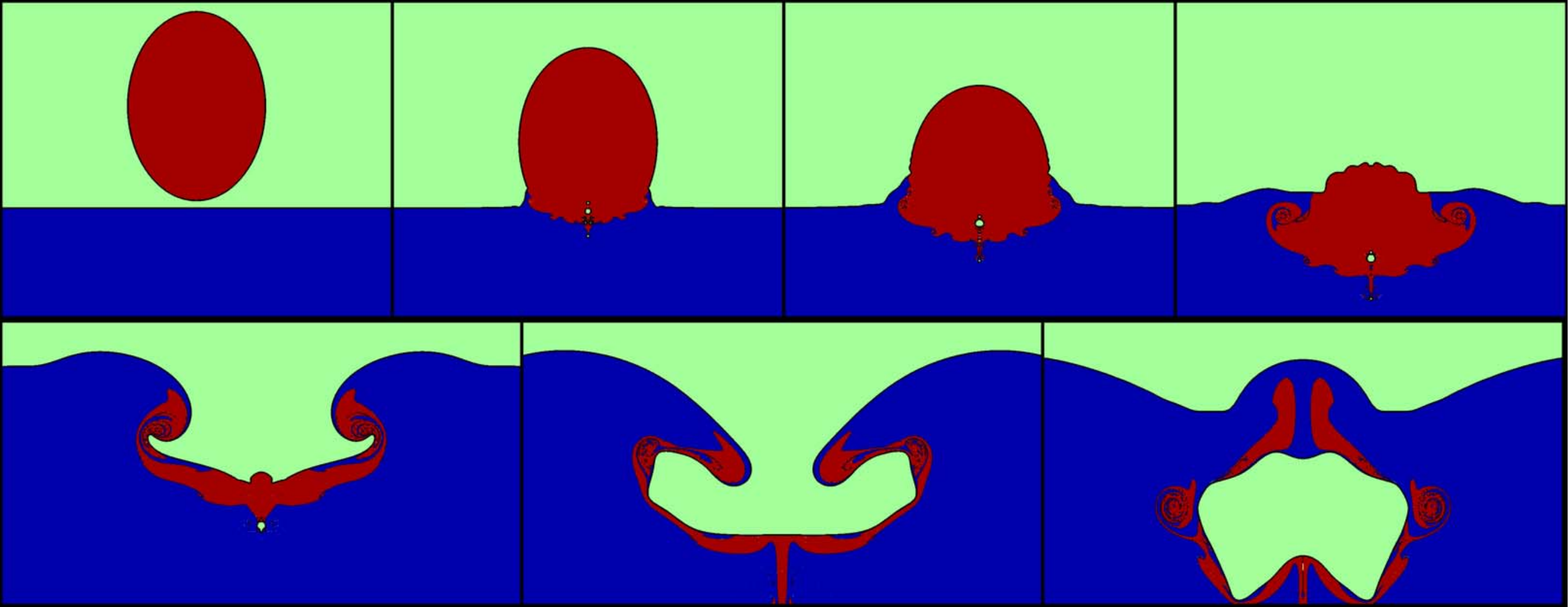}
	\includegraphics[width=0.8\linewidth]{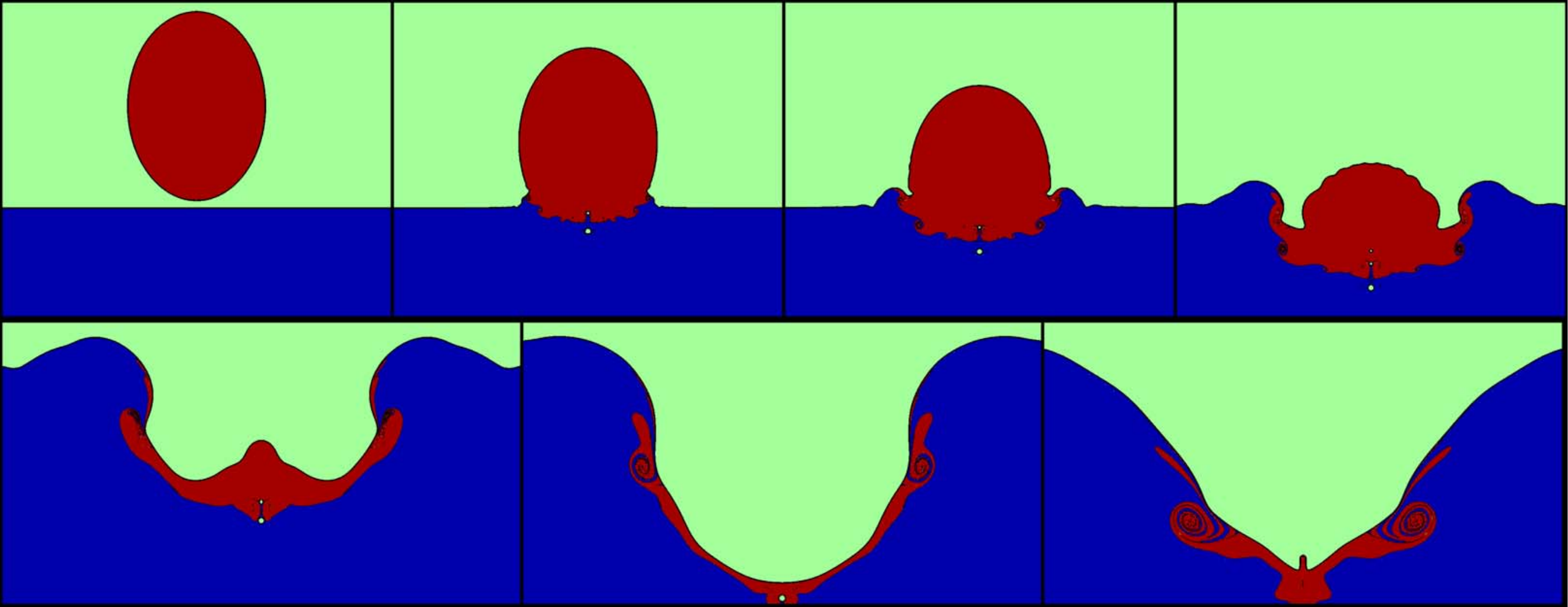}
  \caption{Sequence of simulations at the limit of the large bubble entrapment region ($D =$~5~mm, $V =$~0.9~m/s, $a =$~0.9),
	with maximum level of refinement 10, 11 and 12 from top to bottom.}
  \label{fig:largeBubble-V09}
 \end{center}
\end{figure}

However, this refinement was still not high enough to identify precisely the boundaries of the parameter space.
For $V = 0.9$~m/s for instance, the large bubble entrapment was observed at maximum level of refinement 11 ($ D / \delta x = 532$),
but not for 10 ($ D / \delta x = 266$) or 12 ($ D / \delta x = 1065$) as shown in Fig.~\ref{fig:largeBubble-V09}).

\clearpage
\subsection{Additions on vorticity and kinetic energy}

\begin{figure}[h!]
 \begin{center}
  \includegraphics[width=0.49\linewidth]{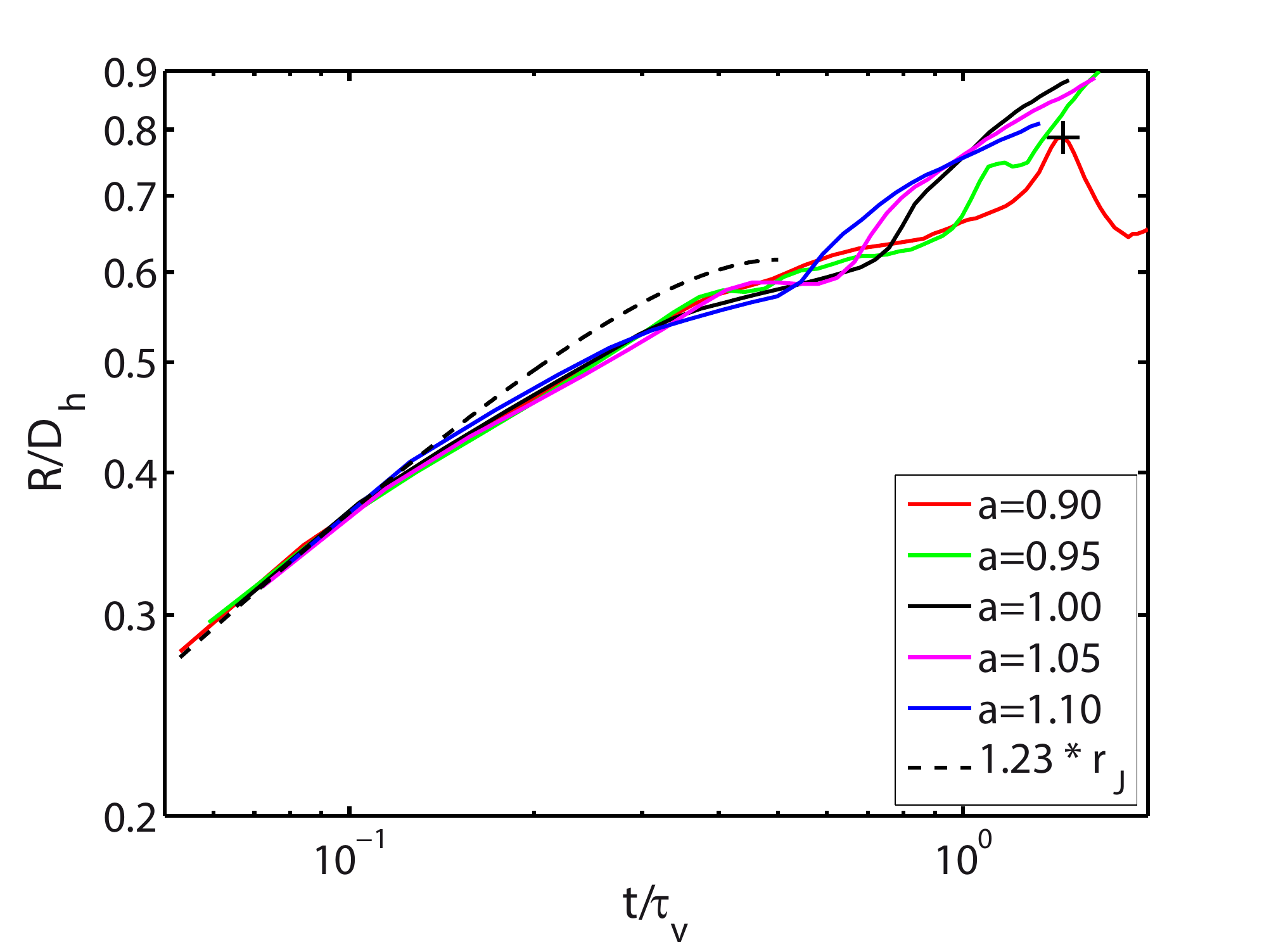}
  \caption{Radial location of the maximum vorticity point in the main vortex shed behind the neck.
	The dashed line is the equation for the position of the neck that was reported in \cite{Thoraval2012}.}
  \label{fig:VorticityRt}
 \end{center}
\end{figure}

This section focuses on the impact conditions of Fig.~\ref{fig:compareTauv} ($D=$~5~mm, $V=$~1~m/s),
and also in Figs~\ref{fig:Vorticity} and \ref{fig:Ek} of the main text.
The radial position of the vortex is controlled by the radial width of the drop,
as the curves collapse for early times $t/\tau_v \leq 0.5$ (Fig.~\ref{fig:VorticityRt}).
Fig.~\ref{fig:Vorticity}(c,d) clearly showed that the main effect of the shape is to delay
the pulling on the interface by the vortex, which is thus occurring at a deeper location.

\begin{figure}
 \begin{center}
  \includegraphics[width=0.49\linewidth]{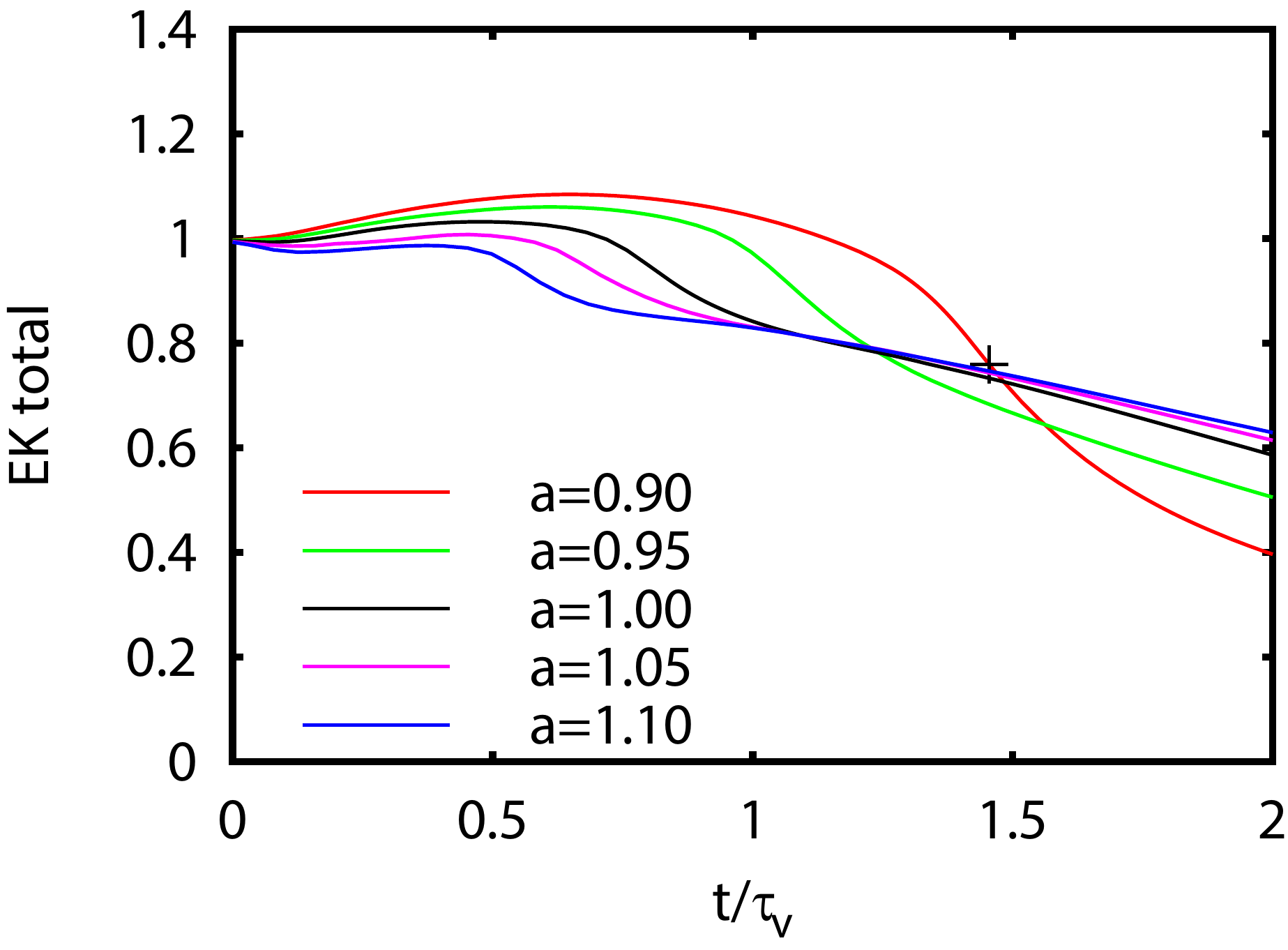}
  \caption{Total kinetic energy.}
  \label{fig:EK}
 \end{center}
\end{figure}

The total kinetic energy can increase slightly at the beginning (Fig.~\ref{fig:EK})
because of the contribution of the surface energy to the radial kinetic energy from pulling of the neck.

\end{document}